\documentclass[aps,pra,reprint,groupedaddress]{revtex4-2}

\usepackage{graphicx}
\usepackage{amsmath}
\usepackage{bm}
\usepackage{braket}
\usepackage{hyperref}
\hypersetup{
    colorlinks=true,
    linkcolor=blue,
    filecolor=blue,      
    urlcolor=blue,
}

\begin{document}

\title{Isolated-core quadrupole excitation of highly excited autoionizing Rydberg states}

\author{M.\,G\'en\'evriez}
\affiliation{Institute of Condensed Matter and Nanosciences, Universit\'e catholique de Louvain, BE-1348 Louvain-la-Neuve, Belgium}
\email[]{matthieu.genevriez@uclouvain.be}
\author{U.\,Eichmann}
\affiliation{Max-Born-Institute, 12489 Berlin, Germany}
\email[]{eichmann@mbi-berlin.de}

\date{\today}

\begin{abstract}
The structure and photoexcitation dynamics of high lying doubly excited states
of the strontium atom with high angular momenta are studied in the vicinity of
the Sr$^+(N=5)$ threshold. The spectra recorded using resonant multiphoton
isolated core excitation are analyzed with calculations based on configuration
interaction with exterior complex scaling, which treats the correlated motion
of the two valence electrons of Sr from first principles. The results are
rationalized with a model based on multichannel quantum-defect theory and
transition dipole moments calculated with a perturbative treatment of electron
correlations. Together, both approaches reveal that most of the lines observed
in the spectra arise from the interaction of a single optically active state,
coupled to the initial state by an electric-dipole transition, with entire
doubly-excited Rydberg series. The long-range electron correlations
responsible for this interaction unexpectedly vanish for identical values of
the initial and final principal quantum numbers, a fact related to the quasi
hydrogenic nature of the high-$l$ Rydberg electron. This special situation,
and in particular the vanishing interaction, leads to the surprising
observation of an electric \emph{quadrupole} isolated-core excitation with a
similar intensity as the neighboring electric dipole transitions.
\end{abstract}

\maketitle

\section{Introduction}
Controlling the quantum numbers of high lying doubly excited (planetary)
states in alkaline earth atoms is possible at an unprecedented degree using
multistep isolated core excitation in combination with the Stark switching
technique~\cite{jones90,eichmann92,camus93}. The excitation scheme has enabled
successful experimental studies on the transition from independent electron
behavior towards strongly correlated electron dynamics in the energy and time
domains~\cite{camus89,eichmann90,pisharody04,zhang13b}. Interpretation of the
complex spectra has been made possible by the development of multichannel
quantum defect theory (MQDT) and $R$-matrix calculations of short range
parameters, which have proven to be very powerful to extract the key
parameters underlying the two-electron dynamics~\cite{gallagher94,aymar96}.
Only recently, calculations based on configuration interaction with exterior
complex scaling (CI-ECS) have been introduced to incorporate the increasing
extent of electron correlation in configuration space as the core is further
excited~\cite{genevriez21}. This not only provides an improved quantitative
description of high lying doubly excited states but also offers a fascinating
visualization of electron correlation in planetary atoms~\cite{genevriez21a}.

In this paper we analyze the spectra of doubly excited Rydberg series of Sr
recorded in the vicinity of the Sr$^+(N=5)$ threshold with CI-ECS. The
theoretical explanation of sections of these spectra, in which an unperturbed
autoionizing Rydberg series attributed to $5gnl$ states~\cite{eichmann03} is
observed despite the absence, at first glance, of dipole coupling to the
initial $5dn_il_i$ state, had initiated a discussion some time ago on how to
apply Fano's lineshape theory in this case~\cite{fano61,eichmann05,cooper05}.
In the present investigation we focus on additional striking features in an
extended set of experimental spectra, which reveal that the problem is even
more complicated than originally discussed. Identifying peculiarities in the
dipole transition elements caused by electron correlations  we  show how the
onset of long-range electronic correlations is responsible for most of the
features in the spectra, and how a quadrupole transition at an optical
wavelength unexpectedly competes on the same ground as dipole allowed
transitions. This latter observation is somewhat unusual since pure quadrupole
transitions have been observed only if  well isolated from dipole allowed
transitions~\cite{margolis03} or via interferences in photoelectron angular
distributions as a result of the breakdown of the dipole
approximation~\cite{martin98,krassig02}. The observation of an isolated-core-quadrupole excitation paves new ways to control and non-destructively detect
Rydberg states by optical manipulation of the ion core~\cite{muni22}.

We report experimental spectra that were recorded from $5d_{5/2}n_i(l_i=12)$
states ($n_i=16-21$) prepared by multiphoton isolated-core excitation to
states in the vicinity of the Sr$^+(5f)$ and Sr$^+(5g)$ ionization thresholds,
as described in Sec.~\ref{sec:experiment}. The large-scale CI-ECS approach
used to calculate and analyze the spectra is presented in
Sec.~\ref{sec:theory}, before comparing and discussing the experimental and
theoretical results in Sec.~\ref{sec:results}. The mechanisms underlying
electric-dipole excitation to states of predominant $5gnl$ character are
elucidated and explained in the light of a simple model, built upon CI-ECS
results, combining a four-channel MQDT approach with a perturbative
calculation of the transition dipole moments (Sec.~\ref{sec:nonICE_E1}). We
analyze the earlier discussion on excitation
mechanisms~\cite{fano61,eichmann05,cooper05} based on the present results in
Sec.~\ref{sec:phase-shifted-continuum}. The presence of electric quadrupole
excitation in the measured the spectra is demonstrated and discussed in
Sec.~\ref{sec:ICE_E2}.

\section{Experiment}\label{sec:experiment}

\begin{figure}
	\includegraphics[width=\columnwidth]{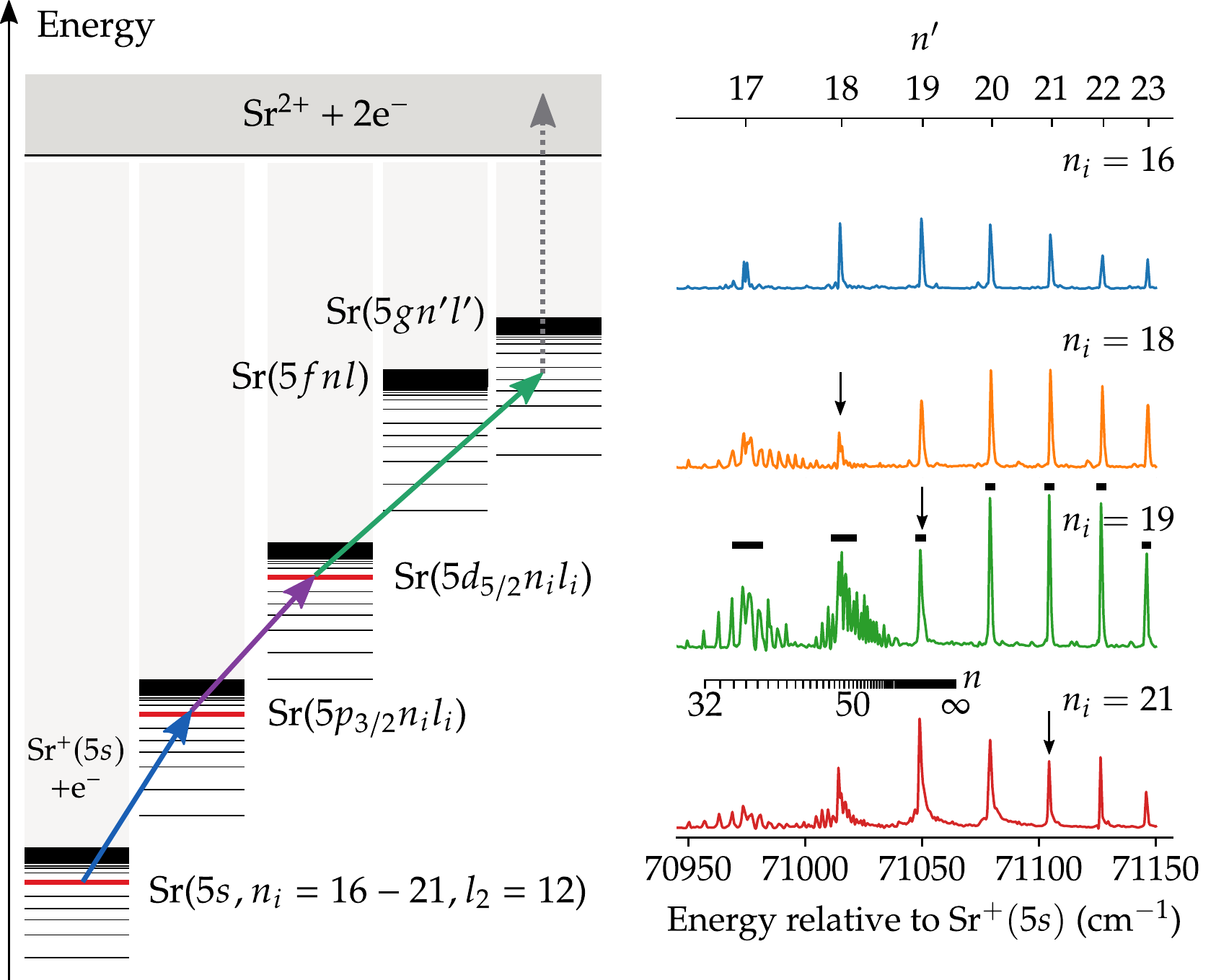}
	\caption{(Left) multiphoton ICE scheme; (Right) photoionization spectra recorded in the region of the Sr$^+(5f)$ and Sr$^+(5g)$ ionization thresholds for $5dn_il_i$ initial states with $l_i \sim 12$ and $n_i=16, 18, 19$ and $21$. The assignment bars on the top axis and lower half of the right panel show the principal quantum numbers $n'$ and $n$ relative to the Sr$^+(5g)$ and Sr$^+(5f)$ thresholds, respectively. The horizontal full black lines in the $n_i=19$ spectrum show the widths of the $5gn'l'$ resonances (see text).}
	\label{fig:experimental_spectra}
\end{figure}

Excitation of the $5fnl$ and $5gn^{\prime }l^{\prime }$ doubly-excited Rydberg
series follows the well established sequential resonant multiphoton excitation
scheme described elsewhere~\cite{Eichmann89,eichmann92,rosen99,eichmann03} and
shown in Fig.~\ref{fig:experimental_spectra}. Note that the
independent-particle quantum numbers adopted for the states classification are
only approximate because of the high degree of electronic correlation. In
brief, Sr atoms in an effusive beam are first excited with two dye lasers
pumped by an excimer laser from the $5s^{2}$ ground state via an intermediate
resonance to a $5sn_ik_i$ Rydberg Stark state in the presence of a ``Stark
switching'' static electric field. When slowly turning off the field, the Stark
state adiabtically evolves into a $5sn_il_i$ Rydberg state with a value of
$l_i$ that is determined by the value of $k_i$ selected upon excitation. The
Stark-switching technique thus allows to prepare the first (outer) electron in
a high and selected angular momentum $l_i$~\cite{freeman76}. Nonadiabatic
effects as the field is switched off and stray electric fields in the chamber
can cause the transfer of a small part of the population to other, neighbouring
$l$ values~\cite{genevriez21a,camus93}.

About 1.5 $\mu $s after the excitation of the outer electron, three more dye
lasers pumped by a second excimer laser interact with the inner electron.
Lasers three and four excite the second (inner) valence electron via the
$5p_{3/2}n_il_i$ to the $5d_{5/2}n_il_i$ resonances (see
Fig.~\ref{fig:experimental_spectra}). A strong fifth laser excites the atom
further to the $5fnl / 5gn^{\prime }l ^{\prime }$ series. The pulse energy of
the fifth laser is typically around 1~mJ while the pulse energies  of the other
dye lasers are kept low ($\sim10~\mu$J). Approximately 200 ns after the fifth
laser pulse, a detection electric-field pulse (20~ns rise time, 12 kV/cm) is applied. As
described in \cite{rosen99}, the detection field together with the photons from
the fifth laser photo/field ionize the $5fnl / 5gn^{\prime }l ^{\prime }$
doubly excited states to produce Sr$^{2+}$ ions. These ions are recorded as a
function of the photon energy of the fifth laser.

The Sr$^{2+}$ spectra recorded from the $5dn_i(l_i \sim 12)$ doubly excited
states with $n_i=16, 18, 19$ and $21$ are shown in the right panel of
Fig.~\ref{fig:experimental_spectra}. The two Rydberg series, converging to the
Sr$^+(5f)$ and Sr$^+(5g)$ ionization thresholds, are labeled by the assignment
bars in the lower and upper parts of the figure, respectively. Although the
spectra appear simple at first, more detailed considerations reveal quite the
opposite and there is in fact some debate on the interpretation of the
underlying excitation
mechanisms~\cite{eichmann03,cooper05,eichmann05,komninos04}. At first glance,
the isolated-core-excitation (ICE) approximation~\cite{cooke78a} does not seem
to work in the last step because the $5d - 5g$ core excitation is forbidden by
electric-dipole selection rules. Consequently, no $5gn'l'$ line should be
observed at all in contrast to the experimental observation. Second, the widths
of the $5gn'l'$ lines abruptly reduce when the photon energy is such that the
initial and final effective principal quantum numbers of the Rydberg electron
are similar, \textit{i.e.}, $n' \simeq n_i$ (vertical arrows in the right panel
of Fig.~\ref{fig:experimental_spectra}). This rapid change is highlighted in
the $n_i=19$ spectrum of Fig.~\ref{fig:experimental_spectra} where the
horizontal full lines show the width of the $5gn'l'$ resonances.

For $n' < n_i=19$ the broad lineshape of the $5gn'l'$ states is discernible
from the enhanced appearance of the $5fnl$ Rydberg series, where $n\gg n'$,
while for $n' \geq 19 $, the spectrum consists only of very narrow autoionizing
resonances apparently no longer strongly interacting with the $5fnl$ states.
When we look at the spectrum $n_i=21$ of Fig.~\ref{fig:experimental_spectra},
we clearly observe strong interaction with the $5fnl$ states for $n' < n_i=
21$, and again, sharp $5gn'l'$ autoionizing resonances  for $n' \geq 21$
indicating strongly reduced interaction. Finally, inspecting  the spectrum
$n_i=16$ of Fig.~\ref{fig:experimental_spectra},  we observe only weakly
interacting $5gn'l'$ states with  $n' > n_i = 16$ throughout the spectrum.
Summarizing the observation we find that the apparent interaction between the
$5fnl$ and $5gn'l'$ series is strong for $n' < n_i$ and weaker for $n' \ge
n_i$. However, the abrupt change of linewidths (or interaction strengths)
cannot be explained by a strong energy dependence of the interaction because,
in this case, the change would be independent of $n_i$.

\section{CI-ECS Theory}\label{sec:theory}

To shed light on the mechanisms responsible for $5dn_il_i - 5gn'l'$ excitation
and to understand the origin of the $n_i$-dependent linewidths, we carried out
calculations using the CI-ECS method~\cite{genevriez21} which treats the
motion of the two valence electrons of Sr from first principles. Such
calculations go beyond other widely used method because they take long-range
electrostatic and exchange interactions between the outer and inner electrons
into account and allow the photoionization cross section to be calculated
without relying on the approximations of the ICE model. This is crucial for
the high lying doubly excited states considered here because electron
correlations are nonnegligible over a large region of configuration space. The
large density of states and the large number of channels of the problem make
the calculations challenging and require the use of well optimized basis
functions and computational methods (see
\cite{genevriez21,genevriez21a,genevriez19b,wehrli19} for details).

\begin{figure}
	\includegraphics[width=\columnwidth]{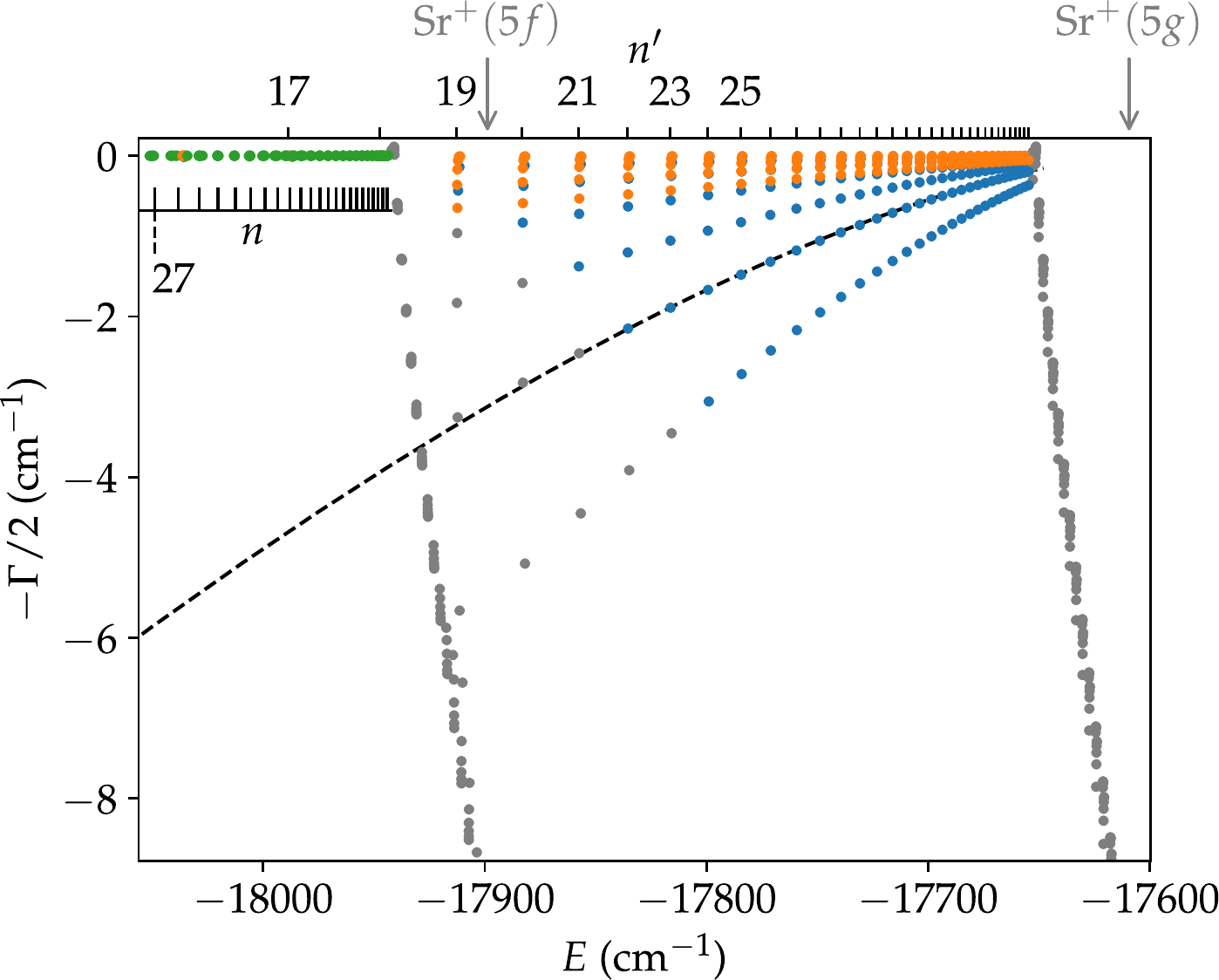}
	\caption{Energies relative to Sr$^{2+}$ and widths of autoionizing states with predominant $5fn(l=12)$, $5gn(l=11)$ and $5gn(l=13)$ character (full green, blue and orange circles, respectively) calculated with CI-ECS for all possible $L$ values. The dashed line shows the $n^{-3}$ scaling of the autoionization rates above the Sr$^{+}(5f)$ threshold. The upper horizontal axis shows the principal quantum numbers $n'$ relative to the $5g$ ionization threshold and the positions of the $5f$ and $5g$ thresholds. The lower assignment bar shows the principal quantum numbers $n$ relative to the $5f$ threshold. The full gray circles are continuum states (see text).}
	\label{fig:eigenvalues}
\end{figure}

Briefly, a two-electron Hamiltonian describing the valence electrons of Sr is
constructed using an empirical model potential for the Sr$^{2+}$ closed-shell
core~\cite{aymar96}. The Hamiltonian matrix is calculated using a basis of
numerical antisymmetrized two-electron functions built from products of two
one-electron spin-orbitals. We use exterior complex
scaling~\cite{simon79,nicolaides1978} to treat the autoionizing resonances and
continuum processes at the heart of the present study. The radial coordinate
of each electron is rotated into the complex plane by an angle
$\theta=5^\circ$ for radial distances greater than $R_0$, with $R_0=150$~$a_0$
chosen to ensure rapid convergence. The one-electron radial functions are
calculated along this complex contour by solving the one-electron
Schr\"odinger equation for the single valence electron of Sr$^+$ with a
finite-element discrete-variable-representation (FEM-DVR)~\cite{rescigno00}.
The calculation parameters and details can be found in
Ref.~\cite{genevriez21a}. In the present work, we used the $LS$
angular-momentum coupling-scheme and did not include spin-orbit interaction as
it is negligible for electrons in the Sr$^+(5f)$, Sr$^+(5g)$ and Rydberg
orbitals. We have checked with a restricted basis set that inclusion of
spin-orbit interaction leaves the spectra unchanged.

The complex-scaled Hamiltonian matrix has eigenvalues given by $E_i -
\text{i}\frac{\Gamma_i}{2}$, where $E_i$ and $\Gamma_i$ are the energies and
widths of the eigenstates, respectively. The eigenvalues of states with
predominant $5fn(l=12)$, $5gn'(l'=11)$ and $5gn'(l'=13)$ character are shown
in Fig.~\ref{fig:eigenvalues} (full green, blue and orange circles,
respectively). The principal quantum numbers of the discrete Rydberg series
converging to the Sr$^+(5f)$ and Sr$^+(5g)$ thresholds are shown in the
assignment bars. Because the calculations are performed in a box of radius
5300~$a_0$, the highest Rydberg state that can be accurately represented is
$n\sim 50$. States with higher principal quantum numbers belong to
quasicontinua that, above the ionization thresholds, become true
(quasidiscretized) ionization continua (full gray circles). As usual in ECS,
the continuum eigenvalues are rotated by approximately $-2\theta$ with respect
to the real axis. Below the $5f$ threshold, the widths of the states are too
small to be visible on the scale of the figure. Their sharp increase
immediately above the $5f$ threshold correlates with the opening of the
$5f\epsilon l$ channels and indicates that autoionization of $5gn'l'$ states
proceeds predominantly into $5f\epsilon l$ continua. The interaction between
the $5fnl$ and $5gn'l'$ channels is thus large, as was already observed in the
experimental spectra (Sec.~\ref{sec:experiment}). Above the
Sr$^+(5f)$ threshold, the widths of the members of each series decay smoothly
as $n^{-3}$ and their energies follow Rydberg's formula.

The photoionization cross section $\sigma$ is calculated from the
complex-scaled two-electron wavefunctions obtained after diagonalization of
the complex-scaled Hamiltonian, as described
in~\cite{genevriez19b,rescigno75}. Calculations are performed for each
possible term of the $5dn_i(l_i=12)$ initial configuration, which is further
assumed to be non-autoionizing. The spectra are then obtained by averaging the
cross sections, by convolving them with a Gaussian function (FWHM of
0.5~cm$^{-1}$) to simulate the effect of the laser bandwidth, by including a
slight saturation effect~\cite{genevriez21a} (see also
Sec.~\ref{fourchannelsection}), and by shifting them horizontally by the
difference ($-20$~cm$^{-1}$) between the experimental Sr$^+(5d_{5/2}-5g)$ and
calculated Sr$^+(5d-5g)$ energy splittings.

\section{Results}\label{sec:results}

\begin{figure*}
	\centering
	\includegraphics[width=\textwidth]{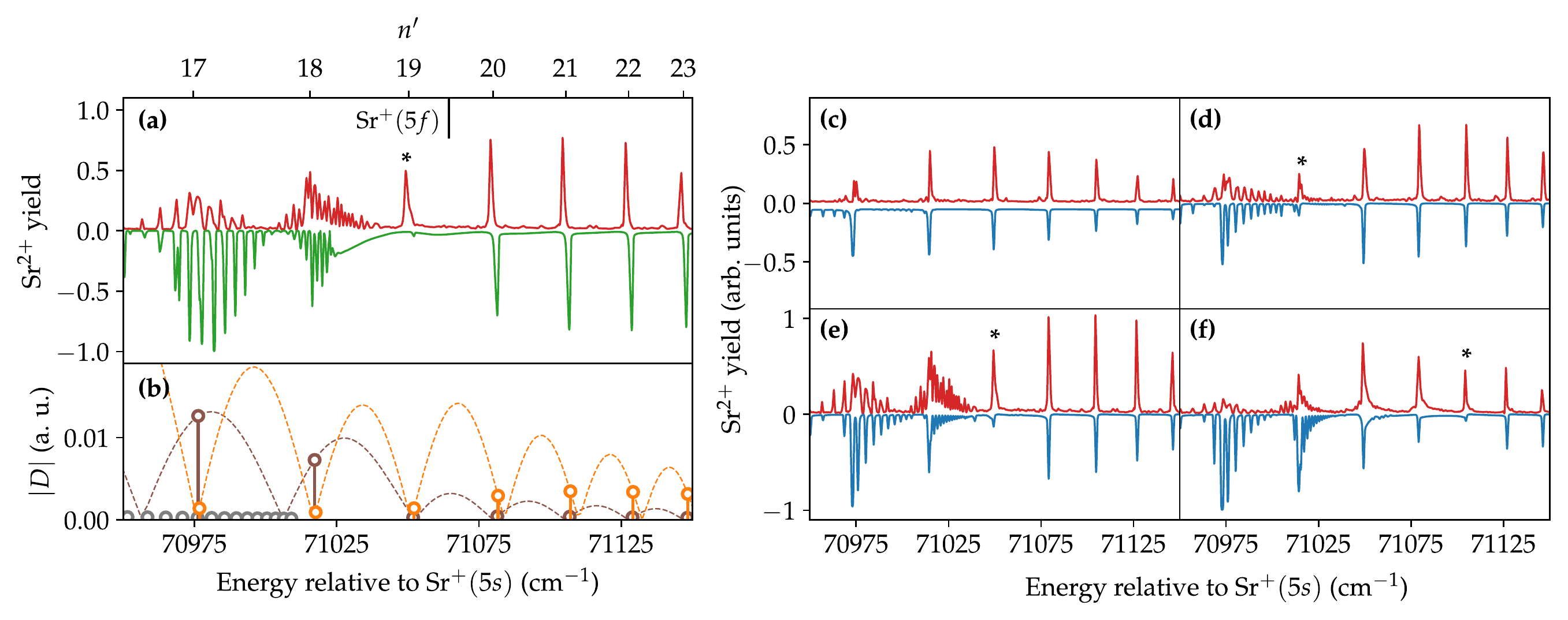}
	\caption{(a) Comparison between the experimental spectrum (red curve) and the spectrum calculated with CI-ECS (green curve) for $n_i=19$ and $l_i=12$. The theoretical spectrum is plotted on a negative scale for clarity. (b) Absolute values of the energy-dependent transition dipole moments $D$ in atomic units (dashed lines) from an initial state with $n_i=19, l_i=12$ to final states with $n'=17-23$ and with $l'=11$ (brown) and $l'=13$ (orange), respectively. Brown and orange circles show the values of $|D|$ at the position of the $5gn'l'$ Rydberg states (assignment bar on the top of panel a) and gray circles show the ICE transition dipole moments to $5fnl$ states with $n=30-45$ and $l=12$ for comparison. (c-f) Comparison of the experimental spectra (red curves) for $n_i=16, 18, 19$ and $21$ (c-f) and the results of a four-channel MQDT analysis. The positions of quadrupole transitions, neglected in the calculations shown in this figure, are marked by asterisks.}
	\label{fig:electricdipole_results}
\end{figure*}

The spectrum calculated within the electric-dipole approximation for an
initial state with $n_i=19$ and $l_i=12$ is shown in
Fig.~\ref{fig:electricdipole_results}(a) and compares well with the
experimental spectrum in terms of both line widths and positions. In
particular, the abrupt reduction of the linewidth for $n'>19$ is reproduced.
The line at $n'=n_i=19$ is, however, not reproduced by the calculation, a point
that is central to the present article and to which we shall come back in
Sec.~\ref{sec:ICE_E2}.

Analysis of the CI-ECS data reveals that, for $n' < n_i$, $5gn'l'$ states with
$l'=l_i-1$ are predominantly excited whereas excitation to $l' = l_i + 1$
states dominates for $n' > n_i$. Because the calculated autoionization rates of
the $l_i-1$ states are larger than those of $l_i+1$ states by typically one
order of magnitude (see Fig.~\ref{fig:eigenvalues}), the $5gn'l'$ lines must
show larger widths for $n' < n_i$ than for $n' > n_i$, as observed in the
experimental and theoretical spectra (see Figs.~\ref{fig:experimental_spectra}
and~\ref{fig:electricdipole_results}). Discrepancies between the theoretical
and experimental line intensities can be attributed to the fact that CI-ECS
calculates the Sr$^+$ ionization rate whereas the experiment records Sr$^{2+}$
ions produced \textit{via} two different photo- and field-ionization mechanisms
with unequal efficiencies below and above the Sr$^+(5f)$
threshold~\cite{rosen99}. This difference is nonetheless insufficient to
explain why the CI-ECS spectra do not reproduce the line in the experimental
spectra at $n'=n_i=19$ (asterisk in Fig.~\ref{fig:electricdipole_results}(a)).

\subsection{non-ICE electric dipole excitation}\label{sec:nonICE_E1}

Before tackling the problem of the vanishing line, we elucidate the mechanism
responsible for excitation from $5dn_il_i$ to $5gn'(l'=l_i\pm1)$ states with
$n_i \neq n'$. A two-channel model based on Fano's treatment of
photoionization~\cite{fano61} was suggested to explain excitation to $5gn'l'$
states~\cite{cooper05,komninos04}. It rests on the breakdown of the
independent-electron approximation and the mixing of $5gn'l'$ states with
predominantly $5fnl$ states by the electron-electron repulsion. By explicitly
calculating the mixing coefficient, we show that this simple ansatz confirms
the vanishing of the line at $n'=n_i$ and, upon inclusion of a third channel,
predicts the change of the linewidths with $n'$ observed in the spectra.
Atomic units are used throughout this section unless stated otherwise.

\subsubsection{Qualitative three-channel analysis}\label{sec:perturbative_dipole}
For the high $l$ values considered here, the long-range dipole part of the
electron-electron repulsion dominates such that $l=l'\pm 1$. Starting from
zeroth-order independent-electron wavefunctions, the mixing coefficient
between $5fnl$ states and $5gn'l'$ states is written within first-order
perturbation theory as
\begin{equation}
	c_{5fnl}^{5gn'l'} = \frac{\braket{5g | r | 5f}\braket{nl | 1/r^2 | n'l'}}{E_{5gn'l'} - E_{5fnl}} \mathcal{B}_{5fnl}^{5gn'l'} ,
	\label{eq:cimix}
\end{equation}
where $\mathcal{B}_{5fnl}^{5gn'l'}$ denotes the integrals over angular coordinates which can be calculated analytically using angular-momentum algebra (see, \textit{e.g.}, Ref.~\cite{cowan81}).
Because the quantum defects of the $5dn_il_i$ and $5fnl$ series extracted from
the CI-ECS data are small ($\delta \lesssim 0.03$), the $\ket{n_il_i}$ and
$\ket{nl_i}$ wavefunctions are to a very good approximation orthogonal. The
photoionization cross section $\sigma$ of a $5dn_il_i$ state \textit{via} the
$5gn'l'$ states can thus be written, within the electric-dipole approximation,
in the simple form
\begin{equation}
	\sigma(\omega)\propto \sum_{l'=l-1}^{l+1}\left|c_{5fn_il_i}^{5gn'l'} \braket{5f | \bm{\hat{\epsilon}}\cdot\bm{r} | 5d}\right|^2 A^2_{5gl'}(\omega) ,
	\label{eq:ice2}
\end{equation}
where $\omega$ is the laser angular frequency. The term $\braket{5f |
\bm{\hat{\epsilon}}\cdot\bm{r} | 5d}$ is the electric-dipole matrix
element of the Sr$^{+}(5d - 5f)$ transition. The density of states $A^2_{5gl'}$
of the channel with orbital angular momentum $l'$ and ion-core state $5g$
consists, to a good approximation, in a series of Lorentzian functions
centered around the positions of the $5gn'l'$ Rydberg states.

Compared to the expression of the cross section derived using the ICE
approximation~\cite{bhatti81}, the transition dipole moment
\begin{equation}
D = c_{5fnl}^{5gn'l'} \braket{5f
	| \bm{\hat{\epsilon}}\cdot\bm{r} | 5d}
\label{dipole}
\end{equation}
  now depends on the photon energy and
on $n_i$ through $n'$,
\begin{equation}
1/2n'^2 = \hbar\omega_{5d5g} + 1/2n_i^2 - \hbar\omega,
\end{equation}
where $\hbar\omega_{5d5g}$ is the Sr$^+(5d-5g)$ energy difference. The
absolute values of the transition dipole moments $D$ from the $5d19(l_i=12)$
state to $5gn'l'$ final states with $l'=11$ and $l'=13$ are shown in
Fig.~\ref{fig:electricdipole_results}(b) as brown and orange circles,
respectively. The dipole moments for $l' = l_i + 1$ and $n' \le n_i$ are very
small and the same is true for $l'=l_i-1$ and $n'\ge n_i$. This occurs
because, in such cases, the dipole moment changes sign near the energies of
the $5gn'l'$ states. The phenomenon, analogous to Cooper minima in
photoionization cross sections~\cite{cooper62a}, traces back to the change of
sign of the $\braket{nl | 1/r^2 | n'l'}$ matrix element in
Eq.~\eqref{eq:cimix}. It is also similar to the predominance, in hydrogenic
oscillator strengths, of $l\to l+1$ transitions for $n_i < n'$ and of $l\to
l-1$ for $n_i > n'$ ~\cite{bethe2012quantum}. For $n'=n_i=19$, both mixing
coefficients $c_{5fnl}^{5gn'l'=l\pm 1}$ are very small and would in fact
vanish for zero quantum defects because the $\braket{nl | 1/r^2 | nl\pm 1}$
matrix element is exactly zero for hydrogenic wavefunctions (see
Ref.~\cite{pasternack62} for a mathematical demonstration).

The three-channel model thus shows that the excitation from $5dn_il_i$ states
to $5gn'l'$ states with $n'\neq n_i$ and $l'=l_i\pm 1$ is enabled by
long-range dipole interactions between the electrons. The conspicuous abrupt
change of interaction strength visible in the four spectra below and above the
$5gn'l'$ states for which $ n'=n_i$ is a consequence of (i) the changing
strengths of the transition dipole moments involving the $5gn'(l_i-1)$ and
$5gn'(l_i+1)$ states (see Fig.~\ref{fig:electricdipole_results}), and (ii) the
vastly different interaction strength of the $5gn'l_i+1$ and $5gn'l_i-1$
series with the $5fnl$ channels. The $5gn'(l_i-1)$ states, which have a much
larger interaction strength with the $5fnl$ channel and thus a much larger
autoionization rate than the $5gn'(l_i+1)$ states, are predominantly excited for $n'
< n_i$, while the weakly interacting $5gn'(l_i+1)$ states, with low
autoionization rates, are favorably excited for $n' > n_i$ (see
Fig.~\ref{fig:eigenvalues}). In essence, excitation of the strongly
interacting states is dominant for $n' < n_i$ while excitation of the weakly
interacting states dominates for $n' > n_i$. The energy- and initial-state
dependence of the transition moments $D$ is thus responsible for many
remarkable properties of the spectra reported here.

\subsubsection{Four-channel quantum defect theory analysis}\label{fourchannelsection}

To confirm the above conclusions quantitatively we use a MQDT approach
\cite{cooke85,woerkom88,giusti84}, where the two-electron wavefunction is
expanded in terms of collision channels
\begin{equation}
\ket{\Psi}= \sum_k A_k \ket{\Xi_k}\ket{\phi_k}.
\end{equation}
Here, $A_k$ is the admixture coefficient of the k$^{th}$ channel, $\Xi_k$
denotes the core electron wave function including  the angular coordinates of
the outer electron (the spin is neglected in our analysis) and $\phi_k $ is the 
radial Coulomb wave function. Following  the  phase-shifted R-matrix  MQDT
approach of Cooke and Cromer \cite{cooke85}, the energy-dependent admixture
coefficients $A_k$ are obtained from
\begin{equation}
	\left[\underline{\mathbf{R}} + \tan (\pi \mathbf{\nu^{(p)}}) \right] \mathbf{a}=0. 
	\label{basicRmatrix}
\end{equation}
The ``phase-shifted" effective quantum numbers $\nu^{(p)}_k$  and admixture
coefficients $a_k^{(p)}$ are defined  by  $\nu^{(p)}_k= \nu_k + \delta_k$ and
$a_k^{(p)}= A_k \cos (\pi \nu^{(p)}_k)$, where $\delta_k$ is the single
channel quantum defect and $ \nu_k$ is the effective quantum number defined by $\nu_k =n - \delta_k$. The phase shifted symmetric matrix
$\underline{\mathbf{R}}$ contains only non-diagonal elements $R_{kj}$ which
describe the coupling strength between channels $k$ and $j$.

To simulate an experimentally observed spectrum we further  need to consider
the relevant channel-excitation dipole moments, which act quasi as filters
weighing the contributions of the energy-dependent channel admixtures (see,
e.g., Fig.~34 of \cite{aymar96}). With energy-dependent collision-channel
dipole moments $D_k$, the total excitation cross section in the case of one
continuum is thus given by
\begin{equation}
 	\sigma \propto (\sum_k A_k D_k)^2,
 	\label{cross}
 \end{equation}
while in the case of many continua we have to take into account the incoherent sum of
the contributions to each continuum (see Eq.~15 of \cite{woerkom88}).

In an extension of the three-channel model discussed qualitatively in
Sec.~\ref{sec:perturbative_dipole}, we are now in a position to simulate the
spectra shown in Fig.~\ref{fig:experimental_spectra}. We use a four-channel
MQDT approach comprising  a common continuum channel, the 5f$n l$  channel and
the 5g$n'( l-1)$ and  5g$n' (l+1)$ channels, denoted by  1 to  4, respectively.
Following Eqs.~45 to 47 of \cite{cooke85} we solved Eq.~\eqref{basicRmatrix}.
Below the Sr$^+(5f)$ ionization limit we have one open and three closed
channels, while above the Sr$^+(5f)$ ionization limit channel 2 becomes open
leaving us with two bound and two open channels. We assume that the spectra are
dominated by exciting the otherwise inherently non-dipole-allowed $5gnl'$
channels through the admixture of the particular $5fn_il$ state. We note that
this admixture  gives rise to the notion that the ionic core is no longer
isolated. The energy-dependent dipole moments $D_3$ and $D_4$ for the different
initial $n_i$ are obtained  from Eq.~\eqref{dipole}. The excitation of the
continuum (channel 1) and of the $5fnl$ channel through the dipole allowed ICE
process of electron shake-up and shake-off is neglected, thus setting
$D_1=D_2=0$. Guided by the CI-ECS calculations we  obtain meaningful numbers
for  the quantum defects and channel interaction strengths. The best match with
the experimental resonance positions is obtained  using $\delta_1 = \delta_2 =0
$ and $\delta_3 = \delta_4 = 0.03$. We note, however, that the energy value  of
the $\text{Sr}^+(5g)$ ionization threshold taken from literature had to be
lowered by 1.8~cm$^{-1}$ to get the best overall agreement. The  different
interaction strengths of the two $5gn'l'$ channels (channels 3,4) with the
$5fnl$ channel (channel 2) are best incorporated using $R_{23}=0.1$ and
$R_{24}=0.4$, which reflects the interaction strengths calculated using the
CI-ECS method. The other channel interaction strengths are set to  $
R_{12}=-0.1$ and $R_{13}=R_{14}=0.1$. The resulting theoretical spectrum is
convolved with an effective laser bandwidth (full width half maximum) of
0.7~cm$^{-1}$. To account for possible saturation effects the measured yield
can be described by  $Y \propto (1-\exp{(-\sigma \Phi)}$~\cite{bhatti83}. The
time integrated laser fluence $\Phi $ in units of photons per unit area is
treated here as fit parameter, which, as it turns out,  only slightly affects
the line intensities in the final results for states lying below the $5g17l'$
states. In all other cases the measured yield is proportional to the calculated
cross section. With this common set of MQDT parameters for the interaction
strengths and quantum defects together with the $n_i$ dependent dipole moments
of Eq.~\eqref{dipole}, we are able to successfully describe the spectra for the
four initial $5dn_il_i$ states, as shown in
Fig.~\ref{fig:electricdipole_results}(c-f). As for CI-ECS, the agreement
between the experimental and theoretical MQDT spectra shown in
Fig.~\ref{fig:electricdipole_results} is excellent, with the exception of the
$n'=n_i$ lines marked by the asterisks.

\subsubsection{Phase-shifted-continuum excitation}\label{sec:phase-shifted-continuum}

In \cite{eichmann03,eichmann05} the idea was put forward that the excitation of
the $5gnl$ resonances above the $\text{Sr}^+(5f)$ limit proceeds entirely
through excitation of a continuum that is locally phase-shifted by its
interaction with the states of the bound series. This does not require
prediagonalization in order to mix the $5fn_il_i$ state into the $5gn'l'$
states that enables excitation (see discussions in \cite{eichmann05,cooper05}).
Within the framework of the four-channel quantum defect analysis of our spectra
presented above, the direct excitation of the $5gn'l'$ channels is no longer
possible but, instead, the dipole moment solely emanates from the transition to
the $5fnl$ channel. To investigate this interpretation of the spectra, we
analyze a simplified quantum-defect-theory model of the $5gnl$ resonances above
the $\text{Sr}^+(5f)$ ionization limit, which involves only two channels for
the sake of clarity. We consider the $5f
\epsilon l $ continuum (open channel $o$) and the $5gnl'$ resonances (bound
channel $b$). The excitation cross section from the initial $5dn_il_i$ is
obtained from Eq.~\eqref{cross} by setting $D_b=0$, since the
excitation of the $5gnl'$ states from the initial $5dn_il_i$ states is dipole
forbidden. The cross section is solely given by the excitation of the
continuum channel
\begin{equation}
	\sigma \propto A_o^2 D_o^2  ,
	\label{ICE}
\end{equation}  
where the dipole moment $D_o$ of the transition is given by
\begin{equation}
	D_o =  \braket{5f
		| \bm{\hat{\epsilon}}\cdot\bm{r} | 5d}
	\braket{\epsilon l
		|n_il_i}
	\propto  D_{ion} \frac{\sin (\pi n_i + \tau)}{E_i-E_f}
	\label{icecont}.
\end{equation}
We defined $D_{ion}=  \braket{5f | \bm{\hat{\epsilon}}\cdot\bm{r} | 5d}$. The
last term is the overlap integral of the initial bound-Rydberg-electron wave
function and the final energy-normalized continuum wave function, describing
the shake-off of the electron into the continuum \cite{story89}. It can be
considered  an extension of the familiar ICE overlap integral, which contains
the initial and final energy-normalized Rydberg electron wavefunction of
autoionizing Rydberg states. $E_i$ and $E_f$ are the excitation energies of the
$5f n_i l_i$ state and of the final $5f\epsilon l_i $ state, respectively. The
negative continuum phase $-\tau$ replaces the quantum defect $ \pi \nu_1$ of
the bound states of the channel. $\tau$ changes by $\pi$ when crossing an
interacting bound state in the continuum ($5gn'l'$), which means that at a
given energy the term $\sin(\pi n_i + \tau)$ becomes equal to $1$ and the
dipole moment becomes large. Following a standard two-channel quantum-defect
approach~\cite{cooke85,giusti84} detailed in
appendix~\ref{sec:mqdt-phase-shifted}, the cross section can be written for
$\delta_i=\delta_o=0$ as

\begin{equation}
	\sigma \propto D_{ion}^2 \frac{R_{ob}^2}{ (E_i-E_f)^2} A_b^2 f(\nu^{(p)}),
	\label{eq:phase-shifted_xsec}
\end{equation}
where $f(\nu^{(p)}) \simeq 1$ in the extended vicinity of each $5gn'l'$
resonance.

As a matter of fact, the cross section in Eq.~\eqref{eq:phase-shifted_xsec}
describing the excitation of $5g$ states seems to be formally identical to the
one obtained from perturbation theory [see Eq.~\eqref{eq:ice2}] if one
associates the term $\frac{R_{ob}^2}{ (E_i-E_f)^2} $ with the square of the
mixing coefficient $c$ given by Eq.~\eqref{eq:cimix}. We note, however, that
in MQDT the interaction strength $V_{nn'}$ between two bound states of
different channels such as a $5fnl$ state and a $5gn'l'$ state  is related to
the channel interaction $R$ by $V_{nn'} =  -R/\pi (n n')^{3/2} $. It scales
smoothly with $n$ and $n'$ and does not account for vanishing interaction for
a particular combination of quantum numbers, as is the case in first-order
perturbation theory where  the $\braket{nl | 1/r^2 | nl'}$ matrix element is
exactly zero for hydrogenic wave functions with $l= l' \pm 1$. Consequently,
the vanishing line obtained in perturbation theory is nonexistent in the MQDT
approach of the present section. Except for this, both approaches seem to be
equivalent and in agreement with experiment, making the use of either approach
and physical interpretation a matter of taste. On the other hand, if one
extends the analysis to take into account two $5gnl$ channels, the striking
change of the dominant excitation to $5gn'(l'=l_i\pm1)$ states below and above
the quantum number of the initial $5dn_il_i$ state will not be present in the
phase-shifted-continuum approach as can be inferred from the spectra displayed
in appendix~\ref{sec:mqdt-phase-shifted}. According to the
phase-shifted-continuum model, both $5gnl'$ channels are excited quasi with
the same strength, i.e., the same dipole matrix element independent of the
initial $5dn_il_i$ state.  Since the full CI-ECS calculations confirm the
existence of the vanishing line and also reproduce the striking change of
excitation for quantum number $n'< n_i$ and $n'> n_i$ , the mixing of
$5fn_il_i$ character into the $5gn'l'$ channels to allow for dipole excitation
seems to be the favored approach.

In essence, the differences between the approaches of
Sec.~\ref{fourchannelsection} and of the present section are due to the choice
of transition dipole moment which, in particular through the vanishing
admixture of $5fn_il_i$ character to $5gn_il'$ states, has a major effect as
it filters some of the transitions out of the spectra. On the contrary, the
vanishing admixture has no consequence on the energies and widths of the
members of the different series, which smoothly evolve with the principal
quantum number (see Fig.~\ref{fig:eigenvalues}) and are well described by
MQDT. In the approach of the present section, the long-range electron
correlations that cause this vanishing were discarded in calculating $D_0$
and, as a result, the vanishing line is not predicted. Such correlations are
naturally included in the CI-ECS approach (see Sec.~\ref{sec:theory}) and can
be incorporated into MQDT by adding a region between the core and exterior
regions where close-coupling equations are numerically solved~\cite{wood1994}.
Alternatively, the model derived in Sec.~\ref{fourchannelsection} proposes an
ad-hoc approach where the effect of long-range electron correlations is
included only where they play a significant role: in the dipole moments. This
way, the spectra are reproduced with a simple set of calculations, albeit much
more limited than the aforementioned approaches.

\subsection{Electric-quadrupole ICE}\label{sec:ICE_E2}

\begin{figure}
	\centering
	\includegraphics[width=\columnwidth]{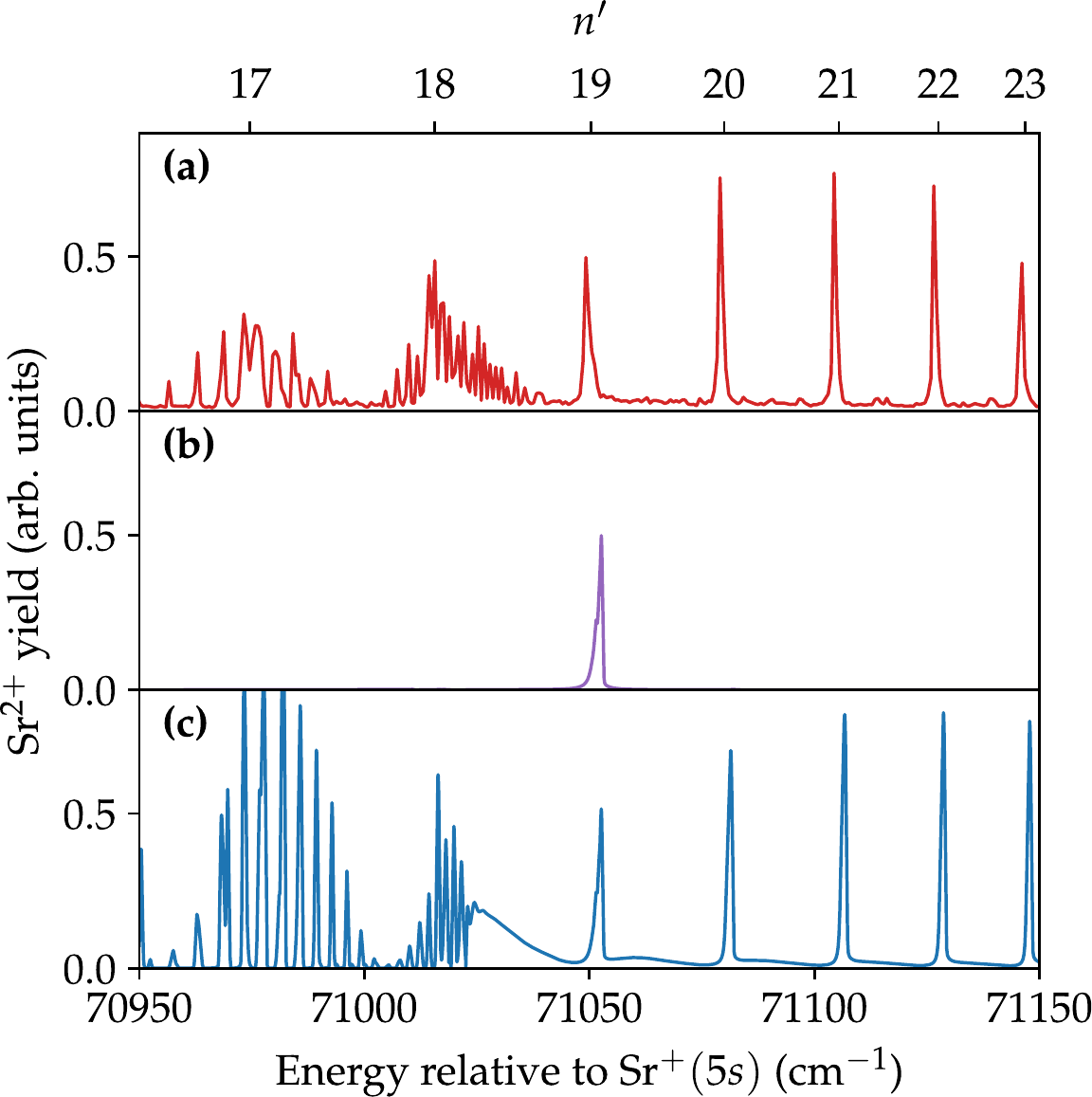}
	\caption{(a) Experimental and (b-c) theoretical CI-ECS spectrum for $n_i=19$ and $l_i=12$. Panel (b) shows the contribution of electric-quadrupole transitions to the theoretical spectrum, and panel (d) shows the total spectrum in which E1 and E2 transitions are summed.}
	\label{fig:th_vs_exp}
\end{figure}

We now consider the $n'=n_i$ lines for which both theoretical calculations
predict weak intensities, in stark contrast to the strong lines observed in
the experiment. We found the calculated intensities to be insensitive to the
values of the principal quantum number, orbital-angular-momentum quantum
number and quantum defect of the Rydberg electron, a fact suggesting that the
discrepancy is not caused by inaccuracies of our wavefunctions but rather by
the omission of an important excitation channel. The spectra in
Fig.~\ref{fig:electricdipole_results} are calculated within the electric
dipole (E1) approximation and excitation occurs \textit{via} the non-ICE
mechanism described earlier. However, it is also possible to directly excite
the Sr$^+$ core from the $5d$ state to the $5g$ state by an electric
\emph{quadrupole} transition (E2). The spectrum resulting from E2 transitions
calculated using CI-ECS is shown in Fig~\ref{fig:th_vs_exp}(b). It displays a
single line around $n'=n_i=19$ with an amplitude similar to the lines in the
E1 spectrum (Fig.~\ref{fig:electricdipole_results}(a)). The existence of a
single line is well described by the independent-electron ICE
model~\cite{cooke78a} because, for the small quantum defects of the high $l$
states under consideration, the overlap integral between the initial and final
Rydberg-state wavefunction entering the cross section is nonnegligible for
$n'=n_i$ only. Combined together, the E1 and E2 spectra are in excellent
agreement with the experimental spectrum (see Fig.~\ref{fig:th_vs_exp}).

The co-existence of electric dipole and quadrupole transitions with similar
magnitudes is unexpected because quadrupole transitions are typically much
weaker~\cite{cowan81}. Two facts are at the origin of this remarkable
situation. First, the Sr$^+(5d - 5g)$ E2 transition is particularly intense
and, for example, the associated Einstein $A$ coefficient calculated with our
one-electron basis functions~\cite{genevriez21a} reaches $388$~s$^{-1}$. In
contrast, the $A$ coefficient of the lower-lying Sr$^+(5s - 4d_{5/2})$
transition is more than a hundred times smaller
($2.559(10)$~s$^{-1}$~\cite{letchumanan05}). Second, the configuration-mixing
coefficients entering the electric-dipole cross section in Eq.~\eqref{eq:ice2}
have values of $\sim 10^{-2}$ and the E1 transitions are thus $10^{4}$ times
less intense than standard E1 transitions.

In cases where the E2 transition is much less intense, one would expect the
absence of a line in the spectra at $n=n_i$. This effect is indeed observed in
the spectra recorded for the Sr($5dn_il_i - 7dn'l'$) transitions with
$l_i=14$~\cite{huang00}. In that case, the coefficients of the configuration
mixing of $7dn'l'$ states with $6fnl$ and $8pnl$ states are $\sim
0.1$~\cite{genevriez21a} whereas the Einstein $A$ coefficient of the Sr$^+(5d -
7d)$ E2 transition is $20$~s$^{-1}$. The Sr($5dn_il_i - 7dn'l'$) transitions
are thus dominated by the E1 contribution~\cite{genevriez21a} and the expected
vanishing of the line at $n'=n_i$ occurs in both experimental and theoretical
spectra~\cite{huang00,genevriez21}.

\section{Conclusion}

We have studied the excitation from $5dn_il_i$ states to doubly excited states
in the vicinity of the Sr$^+(N=5)$ threshold. The experimental spectra show
the excitation of the ion core from the $5d$ to $5g$ orbitals and exhibit an
abrupt change of linewidth at $n'=n_i$. Spectra calculated with CI-ECS fall in
excellent agreement with experimental ones and, together with a simple
four-channel MQDT model built upon the theoretical data, provide a detailed
picture of the complex excitation dynamics of $5dn_il_i$ states. The
electric-dipole excitation to $5gn'l'$ ($n'\neq n_i$) states is made possible
by the admixture of $5fn_il_i$ character due to electron-electron repulsion.
The strong energy dependence of the mixing coefficient, and thus of the
transition dipole moment, favors excitation to $l'=l_i-1$ for $n'<n_i$ and to
$l'=l_i+1$ for $n'>n_i$. Because of the large difference between the rates of
the $l_i-1$ and $l_i+1$ series, the width of the lines in the spectra abruptly
change at the transition between these two regions ($n'=n_i$).

Noticeably, the four-channel model and the CI-ECS calculations in the
electric-dipole approximation both fail to reproduce the strong line at
$n'=n_i$ observed in the experimental spectra. The line is instead attributed
to an electric-quadrupole isolated-core excitation from $5dn_il_i$ states to
$5gn_il_i$ states, which is confirmed by a CI-ECS calculation including E2
contributions. Albeit surprising, the co-existence of E1 and E2 transitions
with the same intensity is explained by the large transition quadrupole moment
of the Sr$^+(5d-5g)$ transition and by the small transition dipole moment due
to the weak admixture of $5fn_il_i$ character into $5gn'l'$ states.

The observation of an electric-quadrupole ICE paves the way for applying new
core-excitation schemes, particularly, to all-optical manipulations of Rydberg
atoms~\cite{teixeira20,burgers22,pham22}. In a recent study, the state of the
Rydberg electron of a Sr atom was detected and coherently manipulated in a
nondestructive manner with a multiphoton electric-dipole ICE
scheme~\cite{muni22}. The scheme relies on a series of four optical
transitions to determine the energy difference $\Delta E$ between the
$4d_{3/2}(|m_j|=3/2)$ and $4d_{3/2}(|m_j|=1/2)$ states of the ion core, which
is due to weak electrostatic interaction between the core- and Rydberg
electrons. Because $\Delta E$ depends on $n$, its measurement served to
determine the principal quantum number of the Rydberg electron in a
non-destructive manner, \textit{i.e.}, without relying on widely used
field-ionization techniques. By demonstrating the possibility to carry out
electric-quadrupole ICE, the present work suggests that the excitation of the ion core
from the $5s_{1/2}$ state to either of the two $4d_{3/2}(|m_j|)$ states can be
used to determine the value $\Delta E$ and thus of $n$ in a nondestructive
manner and with a scheme involving a single photon.

The change of $n$ and $l$ with the non-ICE electric-dipole excitation
discussed above also offers a new route to modify the state of the Rydberg electron
with visible laser light. For $l_i=12$ and zero quantum defects,
the CI mixing coefficients between $4dn'l'$ states and the $5p19l_i$ state
calculated using Eq.~\eqref{eq:cimix} are $\sim 10^{-4}$. Electric dipole
excitation from $5s19l_i$ to $4dn'l'$ states is thus suppressed by $10^{-8}$
compared to Sr$(5sn_il_i - 5pn_il_i)$ excitation, and its intensity is then
comparable to one of the Sr$(5sn_il_i-4dn_il_i)$ E2
transition~\cite{kramida20}. In such a case, dipole and quadrupole transitions
co-exist at a level similar to the one observed in the present spectra. For
larger values of $n$ and $l$, the CI coefficients decrease rapidly and
electric quadrupole transitions dominate. In a Rydberg-atoms quantum simulator, the interaction between neighbouring Rydberg atoms
strongly depends on $n$ and $l$~\cite{weber2017calculation}. The possibility to change $n$ and $l$ with visible light opens the way to changing these interactions in a spatially resolved manner.

\begin{acknowledgments}
	The authors gratefully acknowledge C. Rosen for assistance with the measurements.
\end{acknowledgments}

\bibliographystyle{apsrev4-2}

\begin{thebibliography}{48}%
\makeatletter
\providecommand \@ifxundefined [1]{%
 \@ifx{#1\undefined}
}%
\providecommand \@ifnum [1]{%
 \ifnum #1\expandafter \@firstoftwo
 \else \expandafter \@secondoftwo
 \fi
}%
\providecommand \@ifx [1]{%
 \ifx #1\expandafter \@firstoftwo
 \else \expandafter \@secondoftwo
 \fi
}%
\providecommand \natexlab [1]{#1}%
\providecommand \enquote  [1]{``#1''}%
\providecommand \bibnamefont  [1]{#1}%
\providecommand \bibfnamefont [1]{#1}%
\providecommand \citenamefont [1]{#1}%
\providecommand \href@noop [0]{\@secondoftwo}%
\providecommand \href [0]{\begingroup \@sanitize@url \@href}%
\providecommand \@href[1]{\@@startlink{#1}\@@href}%
\providecommand \@@href[1]{\endgroup#1\@@endlink}%
\providecommand \@sanitize@url [0]{\catcode `\\12\catcode `\$12\catcode
  `\&12\catcode `\#12\catcode `\^12\catcode `\_12\catcode `\%12\relax}%
\providecommand \@@startlink[1]{}%
\providecommand \@@endlink[0]{}%
\providecommand \url  [0]{\begingroup\@sanitize@url \@url }%
\providecommand \@url [1]{\endgroup\@href {#1}{\urlprefix }}%
\providecommand \urlprefix  [0]{URL }%
\providecommand \Eprint [0]{\href }%
\providecommand \doibase [0]{https://doi.org/}%
\providecommand \selectlanguage [0]{\@gobble}%
\providecommand \bibinfo  [0]{\@secondoftwo}%
\providecommand \bibfield  [0]{\@secondoftwo}%
\providecommand \translation [1]{[#1]}%
\providecommand \BibitemOpen [0]{}%
\providecommand \bibitemStop [0]{}%
\providecommand \bibitemNoStop [0]{.\EOS\space}%
\providecommand \EOS [0]{\spacefactor3000\relax}%
\providecommand \BibitemShut  [1]{\csname bibitem#1\endcsname}%
\let\auto@bib@innerbib\@empty
\bibitem [{\citenamefont {Jones}\ and\ \citenamefont
  {Gallagher}(1990)}]{jones90}%
  \BibitemOpen
  \bibfield  {author} {\bibinfo {author} {\bibfnamefont {R.~R.}\ \bibnamefont
  {Jones}}\ and\ \bibinfo {author} {\bibfnamefont {T.~F.}\ \bibnamefont
  {Gallagher}},\ }\href {https://doi.org/10.1103/PhysRevA.42.2655} {\bibfield
  {journal} {\bibinfo  {journal} {Phys. Rev. A}\ }\textbf {\bibinfo {volume}
  {42}},\ \bibinfo {pages} {2655} (\bibinfo {year} {1990})}\BibitemShut
  {NoStop}%
\bibitem [{\citenamefont {Eichmann}\ \emph {et~al.}(1992)\citenamefont
  {Eichmann}, \citenamefont {Lange},\ and\ \citenamefont
  {Sandner}}]{eichmann92}%
  \BibitemOpen
  \bibfield  {author} {\bibinfo {author} {\bibfnamefont {U.}~\bibnamefont
  {Eichmann}}, \bibinfo {author} {\bibfnamefont {V.}~\bibnamefont {Lange}},\
  and\ \bibinfo {author} {\bibfnamefont {W.}~\bibnamefont {Sandner}},\ }\href
  {https://doi.org/10.1103/PhysRevLett.68.21} {\bibfield  {journal} {\bibinfo
  {journal} {Phys. Rev. Lett.}\ }\textbf {\bibinfo {volume} {68}},\ \bibinfo
  {pages} {21} (\bibinfo {year} {1992})}\BibitemShut {NoStop}%
\bibitem [{\citenamefont {Camus}\ \emph {et~al.}(1993)\citenamefont {Camus},
  \citenamefont {Cohen}, \citenamefont {Pruvost},\ and\ \citenamefont
  {Bolovinos}}]{camus93}%
  \BibitemOpen
  \bibfield  {author} {\bibinfo {author} {\bibfnamefont {P.}~\bibnamefont
  {Camus}}, \bibinfo {author} {\bibfnamefont {S.}~\bibnamefont {Cohen}},
  \bibinfo {author} {\bibfnamefont {L.}~\bibnamefont {Pruvost}},\ and\ \bibinfo
  {author} {\bibfnamefont {A.}~\bibnamefont {Bolovinos}},\ }\href
  {https://doi.org/10.1103/PhysRevA.48.R9} {\bibfield  {journal} {\bibinfo
  {journal} {Phys. Rev. A}\ }\textbf {\bibinfo {volume} {48}},\ \bibinfo
  {pages} {R9} (\bibinfo {year} {1993})}\BibitemShut {NoStop}%
\bibitem [{\citenamefont {Camus}\ \emph {et~al.}(1989)\citenamefont {Camus},
  \citenamefont {Gallagher}, \citenamefont {Lecomte}, \citenamefont {Pillet},
  \citenamefont {Pruvost},\ and\ \citenamefont {Boulmer}}]{camus89}%
  \BibitemOpen
  \bibfield  {author} {\bibinfo {author} {\bibfnamefont {P.}~\bibnamefont
  {Camus}}, \bibinfo {author} {\bibfnamefont {T.~F.}\ \bibnamefont
  {Gallagher}}, \bibinfo {author} {\bibfnamefont {J.~M.}\ \bibnamefont
  {Lecomte}}, \bibinfo {author} {\bibfnamefont {P.}~\bibnamefont {Pillet}},
  \bibinfo {author} {\bibfnamefont {L.}~\bibnamefont {Pruvost}},\ and\ \bibinfo
  {author} {\bibfnamefont {J.}~\bibnamefont {Boulmer}},\ }\href
  {https://doi.org/10.1103/PhysRevLett.62.2365} {\bibfield  {journal} {\bibinfo
   {journal} {Phys. Rev. Lett.}\ }\textbf {\bibinfo {volume} {62}},\ \bibinfo
  {pages} {2365} (\bibinfo {year} {1989})}\BibitemShut {NoStop}%
\bibitem [{\citenamefont {Eichmann}\ \emph {et~al.}(1990)\citenamefont
  {Eichmann}, \citenamefont {Lange},\ and\ \citenamefont
  {Sandner}}]{eichmann90}%
  \BibitemOpen
  \bibfield  {author} {\bibinfo {author} {\bibfnamefont {U.}~\bibnamefont
  {Eichmann}}, \bibinfo {author} {\bibfnamefont {V.}~\bibnamefont {Lange}},\
  and\ \bibinfo {author} {\bibfnamefont {W.}~\bibnamefont {Sandner}},\ }\href
  {https://doi.org/10.1103/PhysRevLett.64.274} {\bibfield  {journal} {\bibinfo
  {journal} {Phys. Rev. Lett.}\ }\textbf {\bibinfo {volume} {64}},\ \bibinfo
  {pages} {274} (\bibinfo {year} {1990})}\BibitemShut {NoStop}%
\bibitem [{\citenamefont {Pisharody}\ and\ \citenamefont
  {Jones}(2004)}]{pisharody04}%
  \BibitemOpen
  \bibfield  {author} {\bibinfo {author} {\bibfnamefont {S.~N.}\ \bibnamefont
  {Pisharody}}\ and\ \bibinfo {author} {\bibfnamefont {R.~R.}\ \bibnamefont
  {Jones}},\ }\href {https://doi.org/10.1126/science.1092220} {\bibfield
  {journal} {\bibinfo  {journal} {Science}\ }\textbf {\bibinfo {volume}
  {303}},\ \bibinfo {pages} {813} (\bibinfo {year} {2004})}\BibitemShut
  {NoStop}%
\bibitem [{\citenamefont {Zhang}\ \emph {et~al.}(2013)\citenamefont {Zhang},
  \citenamefont {Jones},\ and\ \citenamefont {Robicheaux}}]{zhang13b}%
  \BibitemOpen
  \bibfield  {author} {\bibinfo {author} {\bibfnamefont {X.}~\bibnamefont
  {Zhang}}, \bibinfo {author} {\bibfnamefont {R.~R.}\ \bibnamefont {Jones}},\
  and\ \bibinfo {author} {\bibfnamefont {F.}~\bibnamefont {Robicheaux}},\
  }\href {https://doi.org/10.1103/PhysRevLett.110.023002} {\bibfield  {journal}
  {\bibinfo  {journal} {Phys. Rev. Lett.}\ }\textbf {\bibinfo {volume} {110}},\
  \bibinfo {pages} {023002} (\bibinfo {year} {2013})}\BibitemShut {NoStop}%
\bibitem [{\citenamefont {Gallagher}(1994)}]{gallagher94}%
  \BibitemOpen
  \bibfield  {author} {\bibinfo {author} {\bibfnamefont {T.~F.}\ \bibnamefont
  {Gallagher}},\ }\href {https://doi.org/10.1017/CBO9780511524530} {\emph
  {\bibinfo {title} {Rydberg {{Atoms}}}}}\ (\bibinfo  {publisher} {{Cambridge
  University Press}},\ \bibinfo {address} {{Cambridge}},\ \bibinfo {year}
  {1994})\BibitemShut {NoStop}%
\bibitem [{\citenamefont {Aymar}\ \emph {et~al.}(1996)\citenamefont {Aymar},
  \citenamefont {Greene},\ and\ \citenamefont {{Luc-Koenig}}}]{aymar96}%
  \BibitemOpen
  \bibfield  {author} {\bibinfo {author} {\bibfnamefont {M.}~\bibnamefont
  {Aymar}}, \bibinfo {author} {\bibfnamefont {C.~H.}\ \bibnamefont {Greene}},\
  and\ \bibinfo {author} {\bibfnamefont {E.}~\bibnamefont {{Luc-Koenig}}},\
  }\href {https://doi.org/10.1103/RevModPhys.68.1015} {\bibfield  {journal}
  {\bibinfo  {journal} {Rev. Mod. Phys.}\ }\textbf {\bibinfo {volume} {68}},\
  \bibinfo {pages} {1015} (\bibinfo {year} {1996})}\BibitemShut {NoStop}%
\bibitem [{\citenamefont {G{\'e}n{\'e}vriez}(2021)}]{genevriez21}%
  \BibitemOpen
  \bibfield  {author} {\bibinfo {author} {\bibfnamefont {M.}~\bibnamefont
  {G{\'e}n{\'e}vriez}},\ }\href {https://doi.org/10.1080/00268976.2020.1861353}
  {\bibfield  {journal} {\bibinfo  {journal} {Mol. Phys.}\ }\textbf {\bibinfo
  {volume} {119}},\ \bibinfo {pages} {e1861353} (\bibinfo {year}
  {2021})}\BibitemShut {NoStop}%
\bibitem [{\citenamefont {G{\'e}n{\'e}vriez}\ \emph {et~al.}(2021)\citenamefont
  {G{\'e}n{\'e}vriez}, \citenamefont {Rosen},\ and\ \citenamefont
  {Eichmann}}]{genevriez21a}%
  \BibitemOpen
  \bibfield  {author} {\bibinfo {author} {\bibfnamefont {M.}~\bibnamefont
  {G{\'e}n{\'e}vriez}}, \bibinfo {author} {\bibfnamefont {C.}~\bibnamefont
  {Rosen}},\ and\ \bibinfo {author} {\bibfnamefont {U.}~\bibnamefont
  {Eichmann}},\ }\href {https://doi.org/10.1103/PhysRevA.104.012812} {\bibfield
   {journal} {\bibinfo  {journal} {Phys. Rev. A}\ }\textbf {\bibinfo {volume}
  {104}},\ \bibinfo {pages} {012812} (\bibinfo {year} {2021})}\BibitemShut
  {NoStop}%
\bibitem [{\citenamefont {Eichmann}\ \emph {et~al.}(2003)\citenamefont
  {Eichmann}, \citenamefont {Gallagher},\ and\ \citenamefont
  {Konik}}]{eichmann03}%
  \BibitemOpen
  \bibfield  {author} {\bibinfo {author} {\bibfnamefont {U.}~\bibnamefont
  {Eichmann}}, \bibinfo {author} {\bibfnamefont {T.~F.}\ \bibnamefont
  {Gallagher}},\ and\ \bibinfo {author} {\bibfnamefont {R.~M.}\ \bibnamefont
  {Konik}},\ }\href {https://doi.org/10.1103/PhysRevLett.90.233004} {\bibfield
  {journal} {\bibinfo  {journal} {Phys. Rev. Lett.}\ }\textbf {\bibinfo
  {volume} {90}},\ \bibinfo {pages} {233004} (\bibinfo {year}
  {2003})}\BibitemShut {NoStop}%
\bibitem [{\citenamefont {Fano}(1961)}]{fano61}%
  \BibitemOpen
  \bibfield  {author} {\bibinfo {author} {\bibfnamefont {U.}~\bibnamefont
  {Fano}},\ }\href {https://doi.org/10.1103/PhysRev.124.1866} {\bibfield
  {journal} {\bibinfo  {journal} {Phys. Rev.}\ }\textbf {\bibinfo {volume}
  {124}},\ \bibinfo {pages} {1866} (\bibinfo {year} {1961})}\BibitemShut
  {NoStop}%
\bibitem [{\citenamefont {Eichmann}\ \emph {et~al.}(2005)\citenamefont
  {Eichmann}, \citenamefont {Gallagher},\ and\ \citenamefont
  {Konik}}]{eichmann05}%
  \BibitemOpen
  \bibfield  {author} {\bibinfo {author} {\bibfnamefont {U.}~\bibnamefont
  {Eichmann}}, \bibinfo {author} {\bibfnamefont {T.~F.}\ \bibnamefont
  {Gallagher}},\ and\ \bibinfo {author} {\bibfnamefont {R.~M.}\ \bibnamefont
  {Konik}},\ }\href {https://doi.org/10.1103/PhysRevLett.94.229302} {\bibfield
  {journal} {\bibinfo  {journal} {Phys. Rev. Lett.}\ }\textbf {\bibinfo
  {volume} {94}},\ \bibinfo {pages} {229302} (\bibinfo {year}
  {2005})}\BibitemShut {NoStop}%
\bibitem [{\citenamefont {Cooper}\ \emph {et~al.}(2005)\citenamefont {Cooper},
  \citenamefont {Greene}, \citenamefont {Langhoff}, \citenamefont {Starace},\
  and\ \citenamefont {Winstead}}]{cooper05}%
  \BibitemOpen
  \bibfield  {author} {\bibinfo {author} {\bibfnamefont {J.~W.}\ \bibnamefont
  {Cooper}}, \bibinfo {author} {\bibfnamefont {C.~H.}\ \bibnamefont {Greene}},
  \bibinfo {author} {\bibfnamefont {P.~W.}\ \bibnamefont {Langhoff}}, \bibinfo
  {author} {\bibfnamefont {A.~F.}\ \bibnamefont {Starace}},\ and\ \bibinfo
  {author} {\bibfnamefont {C.~L.}\ \bibnamefont {Winstead}},\ }\href
  {https://doi.org/10.1103/PhysRevLett.94.229301} {\bibfield  {journal}
  {\bibinfo  {journal} {Phys. Rev. Lett.}\ }\textbf {\bibinfo {volume} {94}},\
  \bibinfo {pages} {229301} (\bibinfo {year} {2005})}\BibitemShut {NoStop}%
\bibitem [{\citenamefont {Margolis}\ \emph {et~al.}(2003)\citenamefont
  {Margolis}, \citenamefont {Huang}, \citenamefont {Barwood}, \citenamefont
  {Lea}, \citenamefont {Klein}, \citenamefont {Rowley}, \citenamefont {Gill},\
  and\ \citenamefont {Windeler}}]{margolis03}%
  \BibitemOpen
  \bibfield  {author} {\bibinfo {author} {\bibfnamefont {H.~S.}\ \bibnamefont
  {Margolis}}, \bibinfo {author} {\bibfnamefont {G.}~\bibnamefont {Huang}},
  \bibinfo {author} {\bibfnamefont {G.~P.}\ \bibnamefont {Barwood}}, \bibinfo
  {author} {\bibfnamefont {S.~N.}\ \bibnamefont {Lea}}, \bibinfo {author}
  {\bibfnamefont {H.~A.}\ \bibnamefont {Klein}}, \bibinfo {author}
  {\bibfnamefont {W.~R.~C.}\ \bibnamefont {Rowley}}, \bibinfo {author}
  {\bibfnamefont {P.}~\bibnamefont {Gill}},\ and\ \bibinfo {author}
  {\bibfnamefont {R.~S.}\ \bibnamefont {Windeler}},\ }\href
  {https://doi.org/10.1103/PhysRevA.67.032501} {\bibfield  {journal} {\bibinfo
  {journal} {Phys. Rev. A}\ }\textbf {\bibinfo {volume} {67}},\ \bibinfo
  {pages} {032501} (\bibinfo {year} {2003})}\BibitemShut {NoStop}%
\bibitem [{\citenamefont {Martin}\ \emph {et~al.}(1998)\citenamefont {Martin},
  \citenamefont {Thompson}, \citenamefont {Bauman}, \citenamefont {Caldwell},
  \citenamefont {Krause}, \citenamefont {Frigo},\ and\ \citenamefont
  {Wilson}}]{martin98}%
  \BibitemOpen
  \bibfield  {author} {\bibinfo {author} {\bibfnamefont {N.~L.~S.}\
  \bibnamefont {Martin}}, \bibinfo {author} {\bibfnamefont {D.~B.}\
  \bibnamefont {Thompson}}, \bibinfo {author} {\bibfnamefont {R.~P.}\
  \bibnamefont {Bauman}}, \bibinfo {author} {\bibfnamefont {C.~D.}\
  \bibnamefont {Caldwell}}, \bibinfo {author} {\bibfnamefont {M.~O.}\
  \bibnamefont {Krause}}, \bibinfo {author} {\bibfnamefont {S.~P.}\
  \bibnamefont {Frigo}},\ and\ \bibinfo {author} {\bibfnamefont
  {M.}~\bibnamefont {Wilson}},\ }\href
  {https://doi.org/10.1103/PhysRevLett.81.1199} {\bibfield  {journal} {\bibinfo
   {journal} {Phys. Rev. Lett.}\ }\textbf {\bibinfo {volume} {81}},\ \bibinfo
  {pages} {1199} (\bibinfo {year} {1998})}\BibitemShut {NoStop}%
\bibitem [{\citenamefont {Kr{\"a}ssig}\ \emph {et~al.}(2002)\citenamefont
  {Kr{\"a}ssig}, \citenamefont {Kanter}, \citenamefont {Southworth},
  \citenamefont {Guillemin}, \citenamefont {Hemmers}, \citenamefont {Lindle},
  \citenamefont {Wehlitz},\ and\ \citenamefont {Martin}}]{krassig02}%
  \BibitemOpen
  \bibfield  {author} {\bibinfo {author} {\bibfnamefont {B.}~\bibnamefont
  {Kr{\"a}ssig}}, \bibinfo {author} {\bibfnamefont {E.~P.}\ \bibnamefont
  {Kanter}}, \bibinfo {author} {\bibfnamefont {S.~H.}\ \bibnamefont
  {Southworth}}, \bibinfo {author} {\bibfnamefont {R.}~\bibnamefont
  {Guillemin}}, \bibinfo {author} {\bibfnamefont {O.}~\bibnamefont {Hemmers}},
  \bibinfo {author} {\bibfnamefont {D.~W.}\ \bibnamefont {Lindle}}, \bibinfo
  {author} {\bibfnamefont {R.}~\bibnamefont {Wehlitz}},\ and\ \bibinfo {author}
  {\bibfnamefont {N.~L.~S.}\ \bibnamefont {Martin}},\ }\href
  {https://doi.org/10.1103/PhysRevLett.88.203002} {\bibfield  {journal}
  {\bibinfo  {journal} {Phys. Rev. Lett.}\ }\textbf {\bibinfo {volume} {88}},\
  \bibinfo {pages} {203002} (\bibinfo {year} {2002})}\BibitemShut {NoStop}%
\bibitem [{\citenamefont {Muni}\ \emph {et~al.}(2022)\citenamefont {Muni},
  \citenamefont {Lachaud}, \citenamefont {Couto}, \citenamefont {Poirier},
  \citenamefont {Teixeira}, \citenamefont {Raimond}, \citenamefont {Brune},\
  and\ \citenamefont {Gleyzes}}]{muni22}%
  \BibitemOpen
  \bibfield  {author} {\bibinfo {author} {\bibfnamefont {A.}~\bibnamefont
  {Muni}}, \bibinfo {author} {\bibfnamefont {L.}~\bibnamefont {Lachaud}},
  \bibinfo {author} {\bibfnamefont {A.}~\bibnamefont {Couto}}, \bibinfo
  {author} {\bibfnamefont {M.}~\bibnamefont {Poirier}}, \bibinfo {author}
  {\bibfnamefont {R.~C.}\ \bibnamefont {Teixeira}}, \bibinfo {author}
  {\bibfnamefont {J.-M.}\ \bibnamefont {Raimond}}, \bibinfo {author}
  {\bibfnamefont {M.}~\bibnamefont {Brune}},\ and\ \bibinfo {author}
  {\bibfnamefont {S.}~\bibnamefont {Gleyzes}},\ }\bibfield  {journal} {\bibinfo
   {journal} {Nat. Phys.}\ }\href {https://doi.org/10.1038/s41567-022-01519-w}
  {10.1038/s41567-022-01519-w} (\bibinfo {year} {2022})\BibitemShut {NoStop}%
\bibitem [{\citenamefont {Eichmann}\ \emph {et~al.}(1989)\citenamefont
  {Eichmann}, \citenamefont {Brockmann}, \citenamefont {Lange},\ and\
  \citenamefont {Sandner}}]{Eichmann89}%
  \BibitemOpen
  \bibfield  {author} {\bibinfo {author} {\bibfnamefont {U.}~\bibnamefont
  {Eichmann}}, \bibinfo {author} {\bibfnamefont {P.}~\bibnamefont {Brockmann}},
  \bibinfo {author} {\bibfnamefont {V.}~\bibnamefont {Lange}},\ and\ \bibinfo
  {author} {\bibfnamefont {W.}~\bibnamefont {Sandner}},\ }\href
  {https://doi.org/10.1088/0953-4075/22/13/003} {\bibfield  {journal} {\bibinfo
   {journal} {Journal of Physics B: Atomic, Molecular and Optical Physics}\
  }\textbf {\bibinfo {volume} {22}},\ \bibinfo {pages} {L361} (\bibinfo {year}
  {1989})}\BibitemShut {NoStop}%
\bibitem [{\citenamefont {Rosen}\ \emph {et~al.}(1999)\citenamefont {Rosen},
  \citenamefont {D{\"o}rr}, \citenamefont {Eichmann},\ and\ \citenamefont
  {Sandner}}]{rosen99}%
  \BibitemOpen
  \bibfield  {author} {\bibinfo {author} {\bibfnamefont {C.}~\bibnamefont
  {Rosen}}, \bibinfo {author} {\bibfnamefont {M.}~\bibnamefont {D{\"o}rr}},
  \bibinfo {author} {\bibfnamefont {U.}~\bibnamefont {Eichmann}},\ and\
  \bibinfo {author} {\bibfnamefont {W.}~\bibnamefont {Sandner}},\ }\href
  {https://doi.org/10.1103/PhysRevLett.83.4514} {\bibfield  {journal} {\bibinfo
   {journal} {Phys. Rev. Lett.}\ }\textbf {\bibinfo {volume} {83}},\ \bibinfo
  {pages} {4514} (\bibinfo {year} {1999})}\BibitemShut {NoStop}%
\bibitem [{\citenamefont {Freeman}\ and\ \citenamefont
  {Kleppner}(1976)}]{freeman76}%
  \BibitemOpen
  \bibfield  {author} {\bibinfo {author} {\bibfnamefont {R.~R.}\ \bibnamefont
  {Freeman}}\ and\ \bibinfo {author} {\bibfnamefont {D.}~\bibnamefont
  {Kleppner}},\ }\href {https://doi.org/10.1103/PhysRevA.14.1614} {\bibfield
  {journal} {\bibinfo  {journal} {Phys. Rev. A}\ }\textbf {\bibinfo {volume}
  {14}},\ \bibinfo {pages} {1614} (\bibinfo {year} {1976})}\BibitemShut
  {NoStop}%
\bibitem [{\citenamefont {Komninos}\ and\ \citenamefont
  {Nicolaides}(2004)}]{komninos04}%
  \BibitemOpen
  \bibfield  {author} {\bibinfo {author} {\bibfnamefont {Y.}~\bibnamefont
  {Komninos}}\ and\ \bibinfo {author} {\bibfnamefont {C.~A.}\ \bibnamefont
  {Nicolaides}},\ }\href {https://doi.org/10.1103/PhysRevA.70.042507}
  {\bibfield  {journal} {\bibinfo  {journal} {Phys. Rev. A}\ }\textbf {\bibinfo
  {volume} {70}},\ \bibinfo {pages} {042507} (\bibinfo {year}
  {2004})}\BibitemShut {NoStop}%
\bibitem [{\citenamefont {Cooke}\ \emph {et~al.}(1978)\citenamefont {Cooke},
  \citenamefont {Gallagher}, \citenamefont {Edelstein},\ and\ \citenamefont
  {Hill}}]{cooke78a}%
  \BibitemOpen
  \bibfield  {author} {\bibinfo {author} {\bibfnamefont {W.~E.}\ \bibnamefont
  {Cooke}}, \bibinfo {author} {\bibfnamefont {T.~F.}\ \bibnamefont
  {Gallagher}}, \bibinfo {author} {\bibfnamefont {S.~A.}\ \bibnamefont
  {Edelstein}},\ and\ \bibinfo {author} {\bibfnamefont {R.~M.}\ \bibnamefont
  {Hill}},\ }\href {https://doi.org/10.1103/PhysRevLett.40.178} {\bibfield
  {journal} {\bibinfo  {journal} {Phys. Rev. Lett.}\ }\textbf {\bibinfo
  {volume} {40}},\ \bibinfo {pages} {178} (\bibinfo {year} {1978})}\BibitemShut
  {NoStop}%
\bibitem [{\citenamefont {G{\'e}n{\'e}vriez}\ \emph {et~al.}(2019)\citenamefont
  {G{\'e}n{\'e}vriez}, \citenamefont {Wehrli},\ and\ \citenamefont
  {Merkt}}]{genevriez19b}%
  \BibitemOpen
  \bibfield  {author} {\bibinfo {author} {\bibfnamefont {M.}~\bibnamefont
  {G{\'e}n{\'e}vriez}}, \bibinfo {author} {\bibfnamefont {D.}~\bibnamefont
  {Wehrli}},\ and\ \bibinfo {author} {\bibfnamefont {F.}~\bibnamefont
  {Merkt}},\ }\href {https://doi.org/10.1103/PhysRevA.100.032517} {\bibfield
  {journal} {\bibinfo  {journal} {Phys. Rev. A}\ }\textbf {\bibinfo {volume}
  {100}},\ \bibinfo {pages} {032517} (\bibinfo {year} {2019})}\BibitemShut
  {NoStop}%
\bibitem [{\citenamefont {Wehrli}\ \emph {et~al.}(2019)\citenamefont {Wehrli},
  \citenamefont {G{\'e}n{\'e}vriez},\ and\ \citenamefont {Merkt}}]{wehrli19}%
  \BibitemOpen
  \bibfield  {author} {\bibinfo {author} {\bibfnamefont {D.}~\bibnamefont
  {Wehrli}}, \bibinfo {author} {\bibfnamefont {M.}~\bibnamefont
  {G{\'e}n{\'e}vriez}},\ and\ \bibinfo {author} {\bibfnamefont
  {F.}~\bibnamefont {Merkt}},\ }\href
  {https://doi.org/10.1103/PhysRevA.100.012515} {\bibfield  {journal} {\bibinfo
   {journal} {Phys. Rev. A}\ }\textbf {\bibinfo {volume} {100}},\ \bibinfo
  {pages} {012515} (\bibinfo {year} {2019})}\BibitemShut {NoStop}%
\bibitem [{\citenamefont {Simon}(1979)}]{simon79}%
  \BibitemOpen
  \bibfield  {author} {\bibinfo {author} {\bibfnamefont {B.}~\bibnamefont
  {Simon}},\ }\href {https://doi.org/10.1016/0375-9601(79)90165-8} {\bibfield
  {journal} {\bibinfo  {journal} {Phys. Lett. A}\ }\textbf {\bibinfo {volume}
  {71}},\ \bibinfo {pages} {211} (\bibinfo {year} {1979})}\BibitemShut
  {NoStop}%
\bibitem [{\citenamefont {Nicolaides}\ and\ \citenamefont
  {Beck}(1978)}]{nicolaides1978}%
  \BibitemOpen
  \bibfield  {author} {\bibinfo {author} {\bibfnamefont {C.~A.}\ \bibnamefont
  {Nicolaides}}\ and\ \bibinfo {author} {\bibfnamefont {D.~R.}\ \bibnamefont
  {Beck}},\ }\href {https://doi.org/10.1016/0375-9601(78)90116-0} {\bibfield
  {journal} {\bibinfo  {journal} {Physics Letters A}\ }\textbf {\bibinfo
  {volume} {65}},\ \bibinfo {pages} {11} (\bibinfo {year} {1978})}\BibitemShut
  {NoStop}%
\bibitem [{\citenamefont {Rescigno}\ and\ \citenamefont
  {McCurdy}(2000)}]{rescigno00}%
  \BibitemOpen
  \bibfield  {author} {\bibinfo {author} {\bibfnamefont {T.~N.}\ \bibnamefont
  {Rescigno}}\ and\ \bibinfo {author} {\bibfnamefont {C.~W.}\ \bibnamefont
  {McCurdy}},\ }\href {https://doi.org/10.1103/PhysRevA.62.032706} {\bibfield
  {journal} {\bibinfo  {journal} {Phys. Rev. A}\ }\textbf {\bibinfo {volume}
  {62}},\ \bibinfo {pages} {032706} (\bibinfo {year} {2000})}\BibitemShut
  {NoStop}%
\bibitem [{\citenamefont {Rescigno}\ and\ \citenamefont
  {McKoy}(1975)}]{rescigno75}%
  \BibitemOpen
  \bibfield  {author} {\bibinfo {author} {\bibfnamefont {T.~N.}\ \bibnamefont
  {Rescigno}}\ and\ \bibinfo {author} {\bibfnamefont {V.}~\bibnamefont
  {McKoy}},\ }\href {https://doi.org/10.1103/PhysRevA.12.522} {\bibfield
  {journal} {\bibinfo  {journal} {Phys. Rev. A}\ }\textbf {\bibinfo {volume}
  {12}},\ \bibinfo {pages} {522} (\bibinfo {year} {1975})}\BibitemShut
  {NoStop}%
\bibitem [{\citenamefont {Cowan}(1981)}]{cowan81}%
  \BibitemOpen
  \bibfield  {author} {\bibinfo {author} {\bibfnamefont {R.~D.}\ \bibnamefont
  {Cowan}},\ }\href@noop {} {\emph {\bibinfo {title} {The Theory of Atomic
  Structure and Spectra}}},\ Los {{Alamos}} Series in Basic and Applied
  Sciences\ (\bibinfo  {publisher} {{University of California Press}},\
  \bibinfo {address} {{Berkeley}},\ \bibinfo {year} {1981})\ Chap.~\bibinfo
  {chapter} {13}\BibitemShut {NoStop}%
\bibitem [{\citenamefont {Bhatti}\ \emph {et~al.}(1981)\citenamefont {Bhatti},
  \citenamefont {Cromer},\ and\ \citenamefont {Cooke}}]{bhatti81}%
  \BibitemOpen
  \bibfield  {author} {\bibinfo {author} {\bibfnamefont {S.~A.}\ \bibnamefont
  {Bhatti}}, \bibinfo {author} {\bibfnamefont {C.~L.}\ \bibnamefont {Cromer}},\
  and\ \bibinfo {author} {\bibfnamefont {W.~E.}\ \bibnamefont {Cooke}},\ }\href
  {https://doi.org/10.1103/PhysRevA.24.161} {\bibfield  {journal} {\bibinfo
  {journal} {Phys. Rev. A}\ }\textbf {\bibinfo {volume} {24}},\ \bibinfo
  {pages} {161} (\bibinfo {year} {1981})}\BibitemShut {NoStop}%
\bibitem [{\citenamefont {Cooper}(1962)}]{cooper62a}%
  \BibitemOpen
  \bibfield  {author} {\bibinfo {author} {\bibfnamefont {J.~W.}\ \bibnamefont
  {Cooper}},\ }\href {https://doi.org/10.1103/PhysRev.128.681} {\bibfield
  {journal} {\bibinfo  {journal} {Phys. Rev.}\ }\textbf {\bibinfo {volume}
  {128}},\ \bibinfo {pages} {681} (\bibinfo {year} {1962})}\BibitemShut
  {NoStop}%
\bibitem [{\citenamefont {Bethe}\ and\ \citenamefont
  {Salpeter}(1977)}]{bethe2012quantum}%
  \BibitemOpen
  \bibfield  {author} {\bibinfo {author} {\bibfnamefont {H.~A.}\ \bibnamefont
  {Bethe}}\ and\ \bibinfo {author} {\bibfnamefont {E.~E.}\ \bibnamefont
  {Salpeter}},\ }\href@noop {} {\emph {\bibinfo {title} {Quantum mechanics of
  one-and two-electron atoms}}}\ (\bibinfo  {publisher} {Plenum Publishing
  Corporation},\ \bibinfo {year} {1977})\BibitemShut {NoStop}%
\bibitem [{\citenamefont {Pasternack}\ and\ \citenamefont
  {Sternheimer}(1962)}]{pasternack62}%
  \BibitemOpen
  \bibfield  {author} {\bibinfo {author} {\bibfnamefont {S.}~\bibnamefont
  {Pasternack}}\ and\ \bibinfo {author} {\bibfnamefont {R.~M.}\ \bibnamefont
  {Sternheimer}},\ }\href {https://doi.org/10.1063/1.1703871} {\bibfield
  {journal} {\bibinfo  {journal} {J. Math. Phys.}\ }\textbf {\bibinfo {volume}
  {3}},\ \bibinfo {pages} {1280} (\bibinfo {year} {1962})}\BibitemShut
  {NoStop}%
\bibitem [{\citenamefont {Cooke}\ and\ \citenamefont {Cromer}(1985)}]{cooke85}%
  \BibitemOpen
  \bibfield  {author} {\bibinfo {author} {\bibfnamefont {W.~E.}\ \bibnamefont
  {Cooke}}\ and\ \bibinfo {author} {\bibfnamefont {C.~L.}\ \bibnamefont
  {Cromer}},\ }\href {https://doi.org/10.1103/PhysRevA.32.2725} {\bibfield
  {journal} {\bibinfo  {journal} {Phys. Rev. A}\ }\textbf {\bibinfo {volume}
  {32}},\ \bibinfo {pages} {2725} (\bibinfo {year} {1985})}\BibitemShut
  {NoStop}%
\bibitem [{\citenamefont {Van~Woerkom}\ and\ \citenamefont
  {Cooke}(1988)}]{woerkom88}%
  \BibitemOpen
  \bibfield  {author} {\bibinfo {author} {\bibfnamefont {L.~D.}\ \bibnamefont
  {Van~Woerkom}}\ and\ \bibinfo {author} {\bibfnamefont {W.~E.}\ \bibnamefont
  {Cooke}},\ }\href {https://doi.org/10.1103/PhysRevA.37.3326} {\bibfield
  {journal} {\bibinfo  {journal} {Phys. Rev. A}\ }\textbf {\bibinfo {volume}
  {37}},\ \bibinfo {pages} {3326} (\bibinfo {year} {1988})}\BibitemShut
  {NoStop}%
\bibitem [{\citenamefont {Giusti-Suzor}\ and\ \citenamefont
  {Fano}(1984)}]{giusti84}%
  \BibitemOpen
  \bibfield  {author} {\bibinfo {author} {\bibfnamefont {A.}~\bibnamefont
  {Giusti-Suzor}}\ and\ \bibinfo {author} {\bibfnamefont {U.}~\bibnamefont
  {Fano}},\ }\href {https://doi.org/10.1088/0022-3700/17/2/008} {\bibfield
  {journal} {\bibinfo  {journal} {Journal of Physics B: Atomic and Molecular
  Physics}\ }\textbf {\bibinfo {volume} {17}},\ \bibinfo {pages} {215}
  (\bibinfo {year} {1984})}\BibitemShut {NoStop}%
\bibitem [{\citenamefont {Bhatti}\ and\ \citenamefont
  {Cooke}(1983)}]{bhatti83}%
  \BibitemOpen
  \bibfield  {author} {\bibinfo {author} {\bibfnamefont {S.~A.}\ \bibnamefont
  {Bhatti}}\ and\ \bibinfo {author} {\bibfnamefont {W.~E.}\ \bibnamefont
  {Cooke}},\ }\href {https://doi.org/10.1103/PhysRevA.28.756} {\bibfield
  {journal} {\bibinfo  {journal} {Phys. Rev. A}\ }\textbf {\bibinfo {volume}
  {28}},\ \bibinfo {pages} {756} (\bibinfo {year} {1983})}\BibitemShut
  {NoStop}%
\bibitem [{\citenamefont {Story}\ and\ \citenamefont {Cooke}(1989)}]{story89}%
  \BibitemOpen
  \bibfield  {author} {\bibinfo {author} {\bibfnamefont {J.~G.}\ \bibnamefont
  {Story}}\ and\ \bibinfo {author} {\bibfnamefont {W.~E.}\ \bibnamefont
  {Cooke}},\ }\href {https://doi.org/10.1103/PhysRevA.39.4610} {\bibfield
  {journal} {\bibinfo  {journal} {Phys. Rev. A}\ }\textbf {\bibinfo {volume}
  {39}},\ \bibinfo {pages} {4610} (\bibinfo {year} {1989})}\BibitemShut
  {NoStop}%
\bibitem [{\citenamefont {Wood}\ and\ \citenamefont {Greene}(1994)}]{wood1994}%
  \BibitemOpen
  \bibfield  {author} {\bibinfo {author} {\bibfnamefont {R.~P.}\ \bibnamefont
  {Wood}}\ and\ \bibinfo {author} {\bibfnamefont {C.~H.}\ \bibnamefont
  {Greene}},\ }\href {https://doi.org/10.1103/PhysRevA.49.1029} {\bibfield
  {journal} {\bibinfo  {journal} {Physical Review A}\ }\textbf {\bibinfo
  {volume} {49}},\ \bibinfo {pages} {1029} (\bibinfo {year}
  {1994})}\BibitemShut {NoStop}%
\bibitem [{\citenamefont {Letchumanan}\ \emph {et~al.}(2005)\citenamefont
  {Letchumanan}, \citenamefont {Wilson}, \citenamefont {Gill},\ and\
  \citenamefont {Sinclair}}]{letchumanan05}%
  \BibitemOpen
  \bibfield  {author} {\bibinfo {author} {\bibfnamefont {V.}~\bibnamefont
  {Letchumanan}}, \bibinfo {author} {\bibfnamefont {M.~A.}\ \bibnamefont
  {Wilson}}, \bibinfo {author} {\bibfnamefont {P.}~\bibnamefont {Gill}},\ and\
  \bibinfo {author} {\bibfnamefont {A.~G.}\ \bibnamefont {Sinclair}},\ }\href
  {https://doi.org/10.1103/PhysRevA.72.012509} {\bibfield  {journal} {\bibinfo
  {journal} {Phys. Rev. A}\ }\textbf {\bibinfo {volume} {72}},\ \bibinfo
  {pages} {012509} (\bibinfo {year} {2005})}\BibitemShut {NoStop}%
\bibitem [{\citenamefont {Huang}\ \emph {et~al.}(2000)\citenamefont {Huang},
  \citenamefont {Rosen}, \citenamefont {Eichmann},\ and\ \citenamefont
  {Sandner}}]{huang00}%
  \BibitemOpen
  \bibfield  {author} {\bibinfo {author} {\bibfnamefont {W.}~\bibnamefont
  {Huang}}, \bibinfo {author} {\bibfnamefont {C.}~\bibnamefont {Rosen}},
  \bibinfo {author} {\bibfnamefont {U.}~\bibnamefont {Eichmann}},\ and\
  \bibinfo {author} {\bibfnamefont {W.}~\bibnamefont {Sandner}},\ }\href
  {https://doi.org/10.1103/PhysRevA.61.040502} {\bibfield  {journal} {\bibinfo
  {journal} {Phys. Rev. A}\ }\textbf {\bibinfo {volume} {61}},\ \bibinfo
  {pages} {040502(R)} (\bibinfo {year} {2000})}\BibitemShut {NoStop}%
\bibitem [{\citenamefont {Teixeira}\ \emph {et~al.}(2020)\citenamefont
  {Teixeira}, \citenamefont {Larrouy}, \citenamefont {Muni}, \citenamefont
  {Lachaud}, \citenamefont {Raimond}, \citenamefont {Gleyzes},\ and\
  \citenamefont {Brune}}]{teixeira20}%
  \BibitemOpen
  \bibfield  {author} {\bibinfo {author} {\bibfnamefont {R.~C.}\ \bibnamefont
  {Teixeira}}, \bibinfo {author} {\bibfnamefont {A.}~\bibnamefont {Larrouy}},
  \bibinfo {author} {\bibfnamefont {A.}~\bibnamefont {Muni}}, \bibinfo {author}
  {\bibfnamefont {L.}~\bibnamefont {Lachaud}}, \bibinfo {author} {\bibfnamefont
  {J.-M.}\ \bibnamefont {Raimond}}, \bibinfo {author} {\bibfnamefont
  {S.}~\bibnamefont {Gleyzes}},\ and\ \bibinfo {author} {\bibfnamefont
  {M.}~\bibnamefont {Brune}},\ }\href
  {https://doi.org/10.1103/PhysRevLett.125.263001} {\bibfield  {journal}
  {\bibinfo  {journal} {Phys. Rev. Lett.}\ }\textbf {\bibinfo {volume} {125}},\
  \bibinfo {pages} {263001} (\bibinfo {year} {2020})}\BibitemShut {NoStop}%
\bibitem [{\citenamefont {Burgers}\ \emph {et~al.}(2022)\citenamefont
  {Burgers}, \citenamefont {Ma}, \citenamefont {Saskin}, \citenamefont
  {Wilson}, \citenamefont {Alarc{\'o}n}, \citenamefont {Greene},\ and\
  \citenamefont {Thompson}}]{burgers22}%
  \BibitemOpen
  \bibfield  {author} {\bibinfo {author} {\bibfnamefont {A.~P.}\ \bibnamefont
  {Burgers}}, \bibinfo {author} {\bibfnamefont {S.}~\bibnamefont {Ma}},
  \bibinfo {author} {\bibfnamefont {S.}~\bibnamefont {Saskin}}, \bibinfo
  {author} {\bibfnamefont {J.}~\bibnamefont {Wilson}}, \bibinfo {author}
  {\bibfnamefont {M.~A.}\ \bibnamefont {Alarc{\'o}n}}, \bibinfo {author}
  {\bibfnamefont {C.~H.}\ \bibnamefont {Greene}},\ and\ \bibinfo {author}
  {\bibfnamefont {J.~D.}\ \bibnamefont {Thompson}},\ }\href
  {https://doi.org/10.1103/PRXQuantum.3.020326} {\bibfield  {journal} {\bibinfo
   {journal} {PRX Quantum}\ }\textbf {\bibinfo {volume} {3}},\ \bibinfo {pages}
  {020326} (\bibinfo {year} {2022})}\BibitemShut {NoStop}%
\bibitem [{\citenamefont {Pham}\ \emph {et~al.}(2022)\citenamefont {Pham},
  \citenamefont {Gallagher}, \citenamefont {Pillet}, \citenamefont {Lepoutre},\
  and\ \citenamefont {Cheinet}}]{pham22}%
  \BibitemOpen
  \bibfield  {author} {\bibinfo {author} {\bibfnamefont {K.-L.}\ \bibnamefont
  {Pham}}, \bibinfo {author} {\bibfnamefont {T.~F.}\ \bibnamefont {Gallagher}},
  \bibinfo {author} {\bibfnamefont {P.}~\bibnamefont {Pillet}}, \bibinfo
  {author} {\bibfnamefont {S.}~\bibnamefont {Lepoutre}},\ and\ \bibinfo
  {author} {\bibfnamefont {P.}~\bibnamefont {Cheinet}},\ }\href
  {https://doi.org/10.1103/PRXQuantum.3.020327} {\bibfield  {journal} {\bibinfo
   {journal} {PRX Quantum}\ }\textbf {\bibinfo {volume} {3}},\ \bibinfo {pages}
  {020327} (\bibinfo {year} {2022})}\BibitemShut {NoStop}%
\bibitem [{\citenamefont {Kramida}\ \emph {et~al.}(2020)\citenamefont
  {Kramida}, \citenamefont {Ralchenko},\ and\ \citenamefont
  {Reader}}]{kramida20}%
  \BibitemOpen
  \bibfield  {author} {\bibinfo {author} {\bibfnamefont {A.}~\bibnamefont
  {Kramida}}, \bibinfo {author} {\bibfnamefont {Y.}~\bibnamefont {Ralchenko}},\
  and\ \bibinfo {author} {\bibfnamefont {J.}~\bibnamefont {Reader}},\ }\href
  {https://doi.org/10.18434/T4W30F} {\bibinfo {title} {{{NIST Atomic Spectra
  Database}} (version 5.8)}} (\bibinfo {year} {2020})\BibitemShut {NoStop}%
\bibitem [{\citenamefont {Weber}\ \emph {et~al.}(2017)\citenamefont {Weber},
  \citenamefont {Tresp}, \citenamefont {Menke}, \citenamefont {Urvoy},
  \citenamefont {Firstenberg}, \citenamefont {B{\"u}chler},\ and\ \citenamefont
  {Hofferberth}}]{weber2017calculation}%
  \BibitemOpen
  \bibfield  {author} {\bibinfo {author} {\bibfnamefont {S.}~\bibnamefont
  {Weber}}, \bibinfo {author} {\bibfnamefont {C.}~\bibnamefont {Tresp}},
  \bibinfo {author} {\bibfnamefont {H.}~\bibnamefont {Menke}}, \bibinfo
  {author} {\bibfnamefont {A.}~\bibnamefont {Urvoy}}, \bibinfo {author}
  {\bibfnamefont {O.}~\bibnamefont {Firstenberg}}, \bibinfo {author}
  {\bibfnamefont {H.~P.}\ \bibnamefont {B{\"u}chler}},\ and\ \bibinfo {author}
  {\bibfnamefont {S.}~\bibnamefont {Hofferberth}},\ }\href
  {https://doi.org/10.1088/1361-6455/aa743a} {\bibfield  {journal} {\bibinfo
  {journal} {J. Phys. B: At. Mol. Opt. Phys.}\ }\textbf {\bibinfo {volume}
  {50}},\ \bibinfo {pages} {133001} (\bibinfo {year} {2017})}\BibitemShut
  {NoStop}%
\end{thebibliography}

\appendix

\section{MQDT treatment of phase-shifted-continuum excitation}\label{sec:mqdt-phase-shifted}

We start from the equation for the cross section given in Sec.~\ref{sec:phase-shifted-continuum},
\begin{equation}
	\sigma \propto A_o^2 D_{ion} \sin (\pi n_i + \tau)/ (E_i-E_f) .
	\label{ice_with_overlap}
\end{equation} 
Following the standard two-channel quantum defect approach\cite{cooke85,giusti84}, $A^2_o=1$ results from normalization  and the continuum phase $\tau$ is given by
\begin{equation}
	\tan(\pi (-\tau +\delta_1)) = R_{ob}^2/\tan \nu^{(p)}_b,
	\label{compatibility}
\end{equation}
where $R_{ob}$ is the  interaction strength between the open and bound channel. Inserting Eq.~\ref{compatibility} into Eq.~\ref{ice_with_overlap} the cross section reads
\begin{equation}
	\sigma \propto \frac{D_{ion}^2}{ (E_i-E_f)^2} \sin^2 \left[ \varphi- \arctan (R_{ob}^2/\tan \nu^{(p)}_b)\right],
\end{equation}
with $\varphi= \pi ( n_i+\delta_o)$. 
Using $\sin y= \frac{\tan y}{\sqrt{1+ \tan^2 y}}$  and addition formulas for sine functions it  can be rewritten as Fano-lineshape-like formula
\begin{equation}
	\sigma \propto \frac{D_{ion}^2 \sin^2 \varphi^2 }{(E_i-E_f)^2} 
	 \frac{\left(q +\epsilon \right)^2}{1+\epsilon^2} 
\end{equation}
where $\epsilon= \tan \nu^{(p)}_b /R_{ob}^2$ and $ q= -1/ \tan \varphi$.
If one takes $n_i=0 $ and $\delta_o=0$ , i.e,   
$\varphi=0$ , we can write 
\begin{equation}
	\sigma \propto D_{ion}^2 \frac{R_{ob}^2}{ (E_i-E_f)^2} \frac{R_{ob}^2}{R_{ob}^4+\tan^2  \nu^{(p)}_b}.
	\label{alternativ approach}
\end{equation}
This can be rewritten to explicitly contain the spectral density $A_b^2$  of bound $5gnl$ autoionizing resonances 
\begin{equation}
	A_b^2= R_{ob}^2 \frac{1+\tan^2  \nu^{(p)}_b}{R_{ob}^4+\tan^2  \nu^{(p)}_b}
\end{equation}
to yield
\begin{equation}
		\sigma \propto D_{ion}^2 \frac{R_{ob}^2}{ (E_i-E_f)^2} A_b^2 f(\nu^{(p)}).
		\label{alternativ approach1}
	\end{equation}
Here,  $f(\nu^{(p)})$ is an  analytical function, which is  $\simeq 1$ in the extended vicinity of each resonance.

\begin{figure}[ht]
	\centering
	\includegraphics[width=0.8\columnwidth]{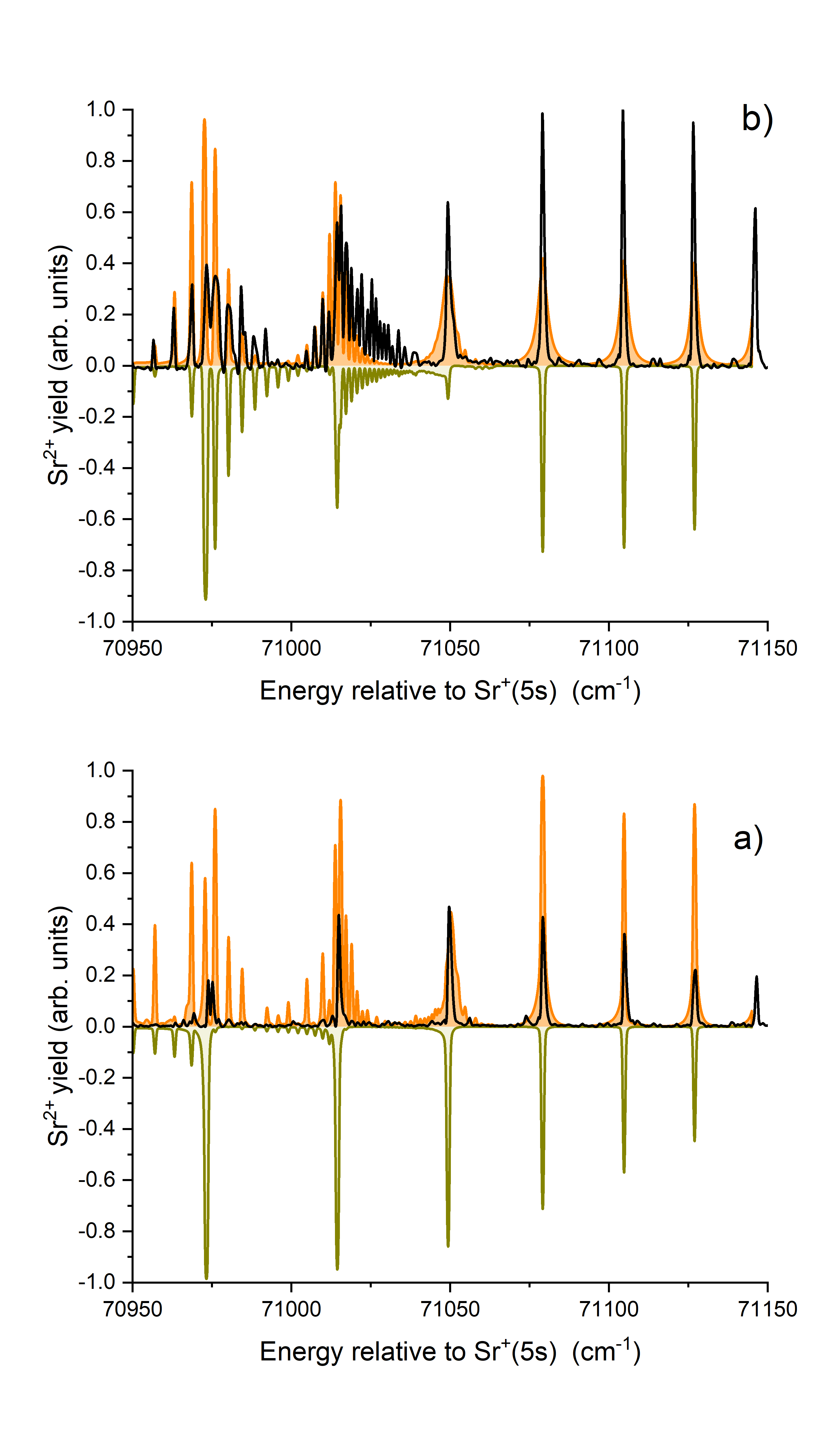}
	\caption{ Spectra of doubly excited states excited  from the   a)   $5d16n_i(l_i=12)$ and b)  $5d19n_i(l_i=12)$ states.  Theoretical curves obtained from a 5-channel analysis  using the MQDT parameters given in section \ref{fourchannelsection}:   Excitation  of the $5f$ channel via ICE ( shake-up and shake off excitation) (orange);   Excitation through the admixture of the $5fn(l=12)$ into the two $5g$ channels (green). Experiment: black curves (same as in Fig.~3). }
	\label{fig:5chan-comp-5f-5g}
\end{figure}

An extension of the model to four channels, as in Sec.~\ref{fourchannelsection} is straightforward. The results of the present approach, based on phase-shifted-continuum excitation, is compared to the mixed perturbative and MQDT approach of Sec.~\ref{fourchannelsection} in Fig.~\ref{fig:5chan-comp-5f-5g}. 

\end{document}